\documentclass[aps, reprint]{revtex4-2}
\usepackage{blindtext}
\usepackage{amssymb}
\usepackage{amsmath}
\usepackage{natbib}

\usepackage{amsfonts}
\usepackage{graphicx}
\usepackage{subfig}
\usepackage{caption}
\usepackage{capt-of}
\usepackage{braket}
\usepackage{array}
\usepackage{float}
\usepackage[colorlinks = true, linkcolor = blue, citecolor = blue, urlcolor = blue]{hyperref}
\newcolumntype{M}[1]{>{\centering\arraybackslash}m{#1}}
\newcolumntype{N}{@{}m{0pt}@{}}
\usepackage{enumerate}
\usepackage{bm}

\begin{document}
\preprint{APS}
\title{Atomic Inversion and Entanglement Dynamics for Squeezed Coherent Thermal States in the Jaynes-Cummings Model}

\author{Koushik Mandal}
\email{ph17d030@smail.iitm.ac.in}
\affiliation{Department of Physics, Indian Institute of Technology Madras, Chennai 600036, India}

\author{M. Venkata Satyanarayana}\thanks{corresponding author}
\email{mvs@iitm.ac.in }
\affiliation{Department of Physics, Indian Institute of Technology Madras, Chennai 600036, India.}

\begin{abstract}
The tussling interplay between the thermal photons and the squeezed photons is discussed. The `classical noise' is represented by the thermal photons and the `quantum noise' is represented by the squeezed photons, which are pitted against each other in the background of a coherent field (represented by the coherent photons). The photon counting distribution (PCD) corresponding to the squeezed coherent thermal states are employed for this purpose. It is  observed that the addition of thermal photons and squeezed photons have counterbalancing effects, by delocalizing and localizing  the PCD, respectively. Various aspects of the  atom-field interaction, like the atomic inversion, and entanglement dynamics in the Jaynes-Cummings model have been investigated. Particular attention is given to the study of atomic inversion and entanglement dynamics due to the  addition of thermal and squeezed photons to the coherent state.  The interplay of thermal photons and squeezed photons have drastic effects on the PCD, atomic inversion and entanglement dynamics of the atom-field interaction.      
\end{abstract}

\maketitle

\section{Introduction}

Mixing of the thermal photons and the squeezed photons to the coherent states have been studied before\cite{ PhysRevA.40.6095,PhysRevA.36.1288,PhysRevA.47.4474, PhysRevA.47.4487, PhysRevA.34.3466, yi1997squeezed, EZAWA1991216, PhysRevA.40.2494}. The photon counting distribution (PCD), the atomic inversion and the entanglement dynamics of these states also have been studied. Satyanarayana \emph{et al}.\cite{satyanarayana1992glauber} studied the PCDs of Glauber-Lachs (GL) states by investigating the effects of addition of thermal noise to  coherent states. It was found that addition of even $2\%$ of the thermal noise makes the peak of the coherent state distribution to fall by $50\%$ of its initial peak. The addition of thermal photons effectively delocalizes the PCD of coherent states. Subeesh \emph{et al}.\cite{subeesh2012effect} observed the effects of squeezing on the PCD of coherent states. It was found that even a small amount of squeezing, e.g., $2\%$ or even $1\%$ of squeezing localizes the PCD of coherent states, i.e., the peaks of distributions upsurge by $100\%$ or more.

In recent years, the squeezed coherent states\cite{RevModPhys.58.1001, PhysRevA.47.5138, PhysRevA.61.010303, PhysRevA.61.022309, PhysRevLett.80.869, PhysRevA.60.937} have been used to study various aspects of quantum optics, quantum information and computation\cite{ hu2013statistical, israel2019entangled, photonics8030072, simidzija2018harvesting, wang2017statistical}. Marian and  Marian have studied the photon counting distributions for the squeezed coherent thermal states (SCTS)\cite{PhysRevA.47.4474, PhysRevA.47.4487}. Very recently, techniques to drive a thermal state into a squeezed thermal state have been discussed in\cite{Dupays2021shortcutstosqueezed}. An SCTS is also a squeezed thermal state in the sense that it is a  displacement operator-shifted  squeezed thermal state in the phase space. Such SCTS have applications in squeezed thermal memory\cite{PhysRevLett.122.040602}. The role of thermal photons vs. squeezed photons is of importance, and we investigate the effects of adding thermal photons to squeezed coherent states and squeezed photons to coherent thermal states. The coherent state is essentially providing a background distribution to bring out the effects of thermal photons vs. squeezed photons. We also focus on  the localization of the PCD  due to the addition of squeezed photons to coherent states and following this, how the addition of thermal photons delocalizes the PCD and almost restores the initial PCD and vice versa. The main objective of this paper is to  bring to light this interesting give and take between squeezed photons and thermal photons, hitherto unnoticed, and  investigate the accompanying effects on the atomic inversion and the entanglement dynamics of the atom-field interaction.  We also study the  effects on the PCD due to the interplay between the  thermal and squeezed photons in the case of coherent squeezed thermal states (CSTS) and study the corresponding effects on the atomic inversion and the entanglement dynamics of the Jaynes-Cummings interaction.

\section{Photon counting distribution} 
The density operator for squeezed coherent thermal states (SCTS) is defined as\cite{PhysRevA.47.4474, PhysRevA.47.4487, yi1997squeezed}

\begin{equation}
\hat{\rho}_{\text{SCT}} = \hat{D}(\alpha)\hat{S}(\zeta)\hat{\rho}_{\text{th}}\hat{S}^{\dagger}(\zeta)\hat{D}^{\dagger}(\alpha),
\end{equation}
where
\begin{equation}
\hat{D}(\alpha) = \exp(\alpha \hat{a}^{\dagger} - \alpha^{*} \hat{a})
\end{equation}
is the displacement operator, for $\alpha$  a complex parameter, and
\begin{equation} 
\hat{S}(\zeta) = \exp\left(-\frac{1}{2}\zeta \hat{a}^{\dagger2} + \frac{1}{2} \zeta^{*}\hat{a}^{2}\right)
\end{equation}
is the squeezing operator with $\zeta = r e^{i\varphi}$. The density operator of a thermal radiation field with a heat bath at temperature $T$ can be written as 

\begin{equation}
\hat{\rho}_{\text{th}} = \frac{1}{1 + N_{th}}\sum_{n=0}^{\infty}\left(
\frac{N_{th}}{N_{th} + 1}\right)^{n}\ket{n}\bra{n},
\end{equation}
where $N_{th}$ is the average number of thermal photons and it is given by
\begin{equation}
    N_{th} = \frac{1}{\exp\left(\frac{h \nu}{k_B T}\right)-1},
\end{equation}
and $k_B$ is the Boltzmann constant.

\begin{figure}
\includegraphics[width = \linewidth]{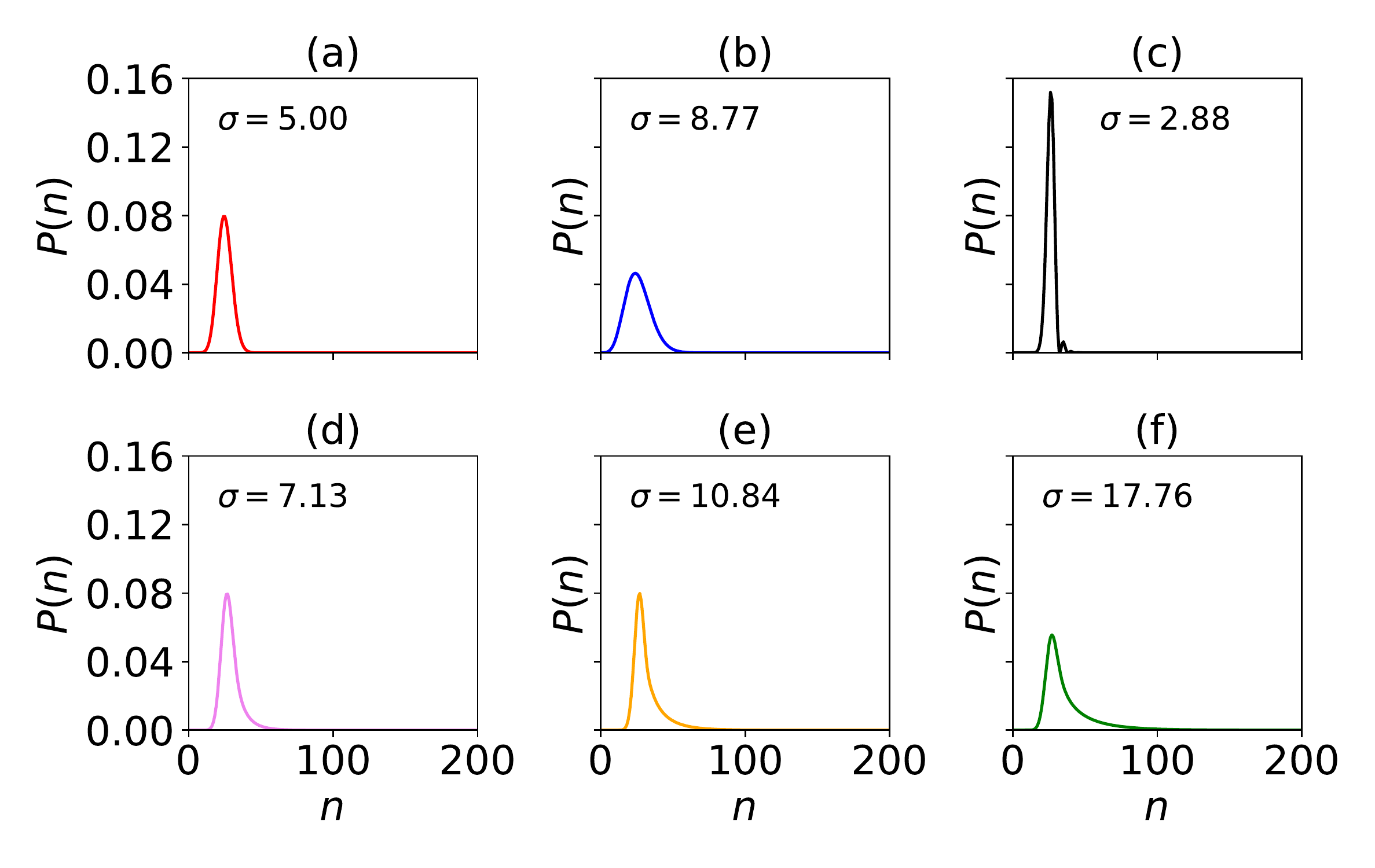}
\caption{Photon counting distributions for SCTS for $(N_{C}, N_{S}, N_{th})$ = (a) $(25, 0, 0)$, (b) $(25, 0, 1)$, (c) $(25, 1, 0)$, (d) $(25, 1, 1)$, (e) $(25, 2, 1)$, (f) $(25, 2, 2)$.}
\label{fig1}
\end{figure}

\begin{figure}
\includegraphics[width = \linewidth]{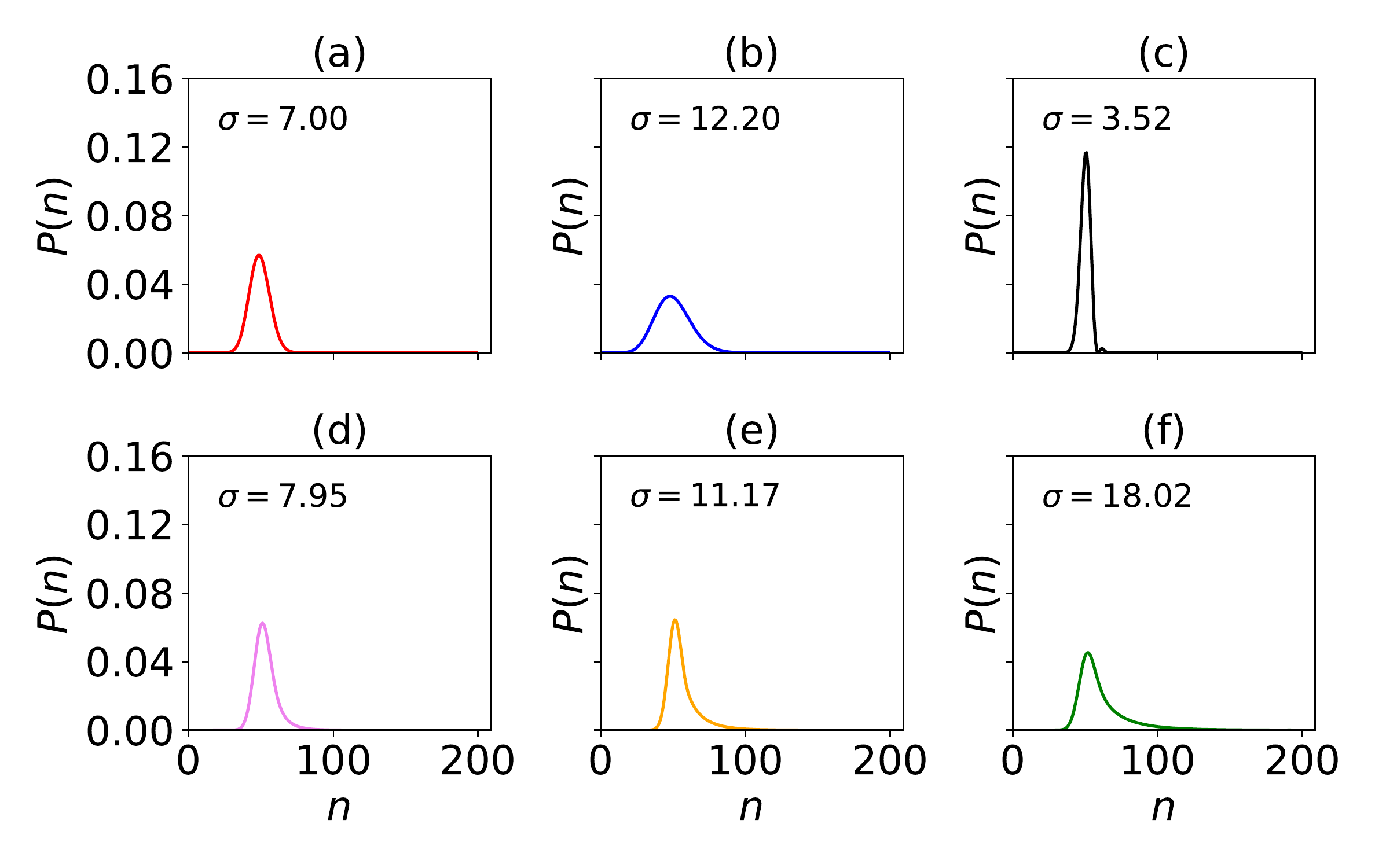}
\caption{Photon counting distributions for SCTS for $(N_{C}, N_{S}, N_{th})$ = (a) $(49, 0, 0)$, (b) $(49, 0, 1)$, (c) $(49, 1, 0)$, (d) $(49, 1, 1)$, (e) $(49, 2, 1)$, (f) $(49, 2, 2)$.}
\label{fig2}
\end{figure}

The analytic expression for the PCD of SCTS can be written as\cite{PhysRevA.47.4474, PhysRevA.47.4487}

\begin{align}
P(l) =& \bra{l}\hat{\rho}_{\text{SCT}}\ket{l}\\
 =&~ \pi Q(0) \tilde{A}^{l}\sum_{q=0}^{l}\frac{1}{q!}\left(\frac{l}{q}\right)\Big|\frac{|\tilde{B}|}{2 \tilde{A}}\Big|^{q}\nonumber\\
&\times \big|H_{q}((2B)^{-1\slash 2} \tilde{C})\big|^{2},
\end{align}

where $\pi Q(0) = R(0,0)$; $R$ is Glauber's $R$-function\cite{PhysRev.131.2766},

\begin{align}
R(0,0) =& \left[(1 + A)^{2} - |B|^{2}\right]^{-1\slash 2} \times\nonumber\\
 &\exp\left\{- \frac{(1+A)|C|^{2} + \frac{1}{2}[B(C^{*})^{2} + B^{*}C^{2}]}{(1+A)^{2}-|B|^{2}}\right\},
\end{align}
where

\begin{align}
A &= N_{th} + (2 N_{th} + 1) (\sinh r)^{2}, \\
B &= - (2 N_{th} + 1)e^{i\varphi} \sinh r \cosh r,\\
C &= \alpha  \hspace{0.5cm} \text{(for SCTS)},\\
C &= \alpha \cosh{r} + \alpha^{*}e^{i \varphi} \sinh{r},\hspace{0.5cm} \text{(for CSTS)}
\end{align}
and
\begin{align}
\tilde{A} &= \frac{A(1 + A) - |B|^{2}}{(1 + A)^{2} - |B|^{2}}, \\
\tilde{B} &= \frac{B}{(1 + A)^{2} - |B|^{2}},\\
\tilde{C} &= \frac{(1 + A)C + BC^{*}}{(1 + A)^{2} - |B|^{2}}.
\end{align}
If we write $\tilde{A}$, $\tilde{B}$ and $\tilde{C}$ in terms of $N_{th}$ and $r$, we get
\begin{align}
\tilde{A} &= \frac{N_{th}(N_{th} + 1)}{N_{th}^{2} + (N_{th} + \frac{1}{2})[1 + \cosh (2r)]}\\
\tilde{B} &= -\frac{e^{i\varphi}(N_{th} + \frac{1}{2})\sinh(2r)}{N_{th}^{2} + (N_{th} + \frac{1}{2})[1 + \cosh (2r)]}\\
\tilde{C} &= \frac{C[\frac{1}{2} + (N_{th}+ \frac{1}{2})\cosh r] - C^{*}e^{i \varphi}(N_{th}+ \frac{1}{2}) \sinh 2r}{N_{th}^{2} + (N_{th} + \frac{1}{2})[1 + \cosh (2r)]} \nonumber\\
\end{align}
and $H_{q}$ is the Hermite polynomial. It is defined as
\begin{equation}
H_{q}(x) = \sum_{j=0}^{\lfloor\frac{ q}{2}\rfloor}\frac{(-1)^{j}q!}{j!(q-2j)!}(2x)^{q-2j}.
\end{equation}
In view of the fact that $P(n)$ in one extreme limit describes a coherent state and in another limit describes a pure squeezed state and in another extreme limit describes a thermal state, the study of atomic inversion of a two-level atom interacting with a single mode of electromagnetic field in a SCTS enables us to observe how the revivals of atomic inversion in a squeezed coherent state develop from ringing revivals\cite{satyanarayana1989ringing} into chaotic oscillations corresponding to a thermal state and vice versa. In fact, various interesting interpolatory paths to study the atomic inversion routes are possible. Also, this approach is useful to study the roles of thermal photons, coherent photons and squeezing on the entanglement dynamics.
	
The photon counting distributions for the SCTS are plotted in Figs. 1 and 2 for different values of average  coherent($N_{C} = |\alpha|^2$), squeezed($N_{S} = \sinh^{2} r$) and thermal($N_{th}$) photons. Fig. 1(a) gives the PCD for $N_{C} = 25$, $N_{S}=0$ and $N_{th} = 0$, which represents a coherent state. In Fig. 1(b), $N_{C} = 25$,  $N_{S} = 0$ and $N_{th}=1$;  the peak of the PCD comes down to half of the initial peak. This reduction in the peak of $P(n)$ was observed by Satyanarayana \textit{et al.}\cite{satyanarayana1992glauber}. In Fig. 1(c), $N_{C} = 25$, $N_{S}=1$ and $N_{th}=0$; the peak increases by almost $100{\%}$ of its initial peak. It is known as the localization of the PCD. This interesting behaviour of the PCD was investigated by Subeesh \textit{et al}.\cite{subeesh2012effect}. In Fig. 1(d), $N_{C} = 25$, $N_s = 1$ and $N_{th} = 1$, i.e., either $N_{th}=1$ is added to the squeezed coherent state corresponding to $N_{C}=25$ and $N_{S}=1$ or $N_{S}=1$ is added to the Glauber-Lachs state (a thermal coherent state) corresponding to $N_{C}=25$ and $N_{th}=1$. The resulting  PCD for the SCTS (corresponding to  $N_{C} = 25$, $N_{S}=1$ and $N_{th} = 1$) is  almost like the PCD of the coherent state (correspond to $N_{C} = 25$, $N_{S}=0$ and $N_{th} = 0$) , but for a little tail. This means the localizing effect of squeezing on a coherent state  is almost compensated by the increase in $N_{th}$. This also means, when the PCD of a coherent state is delocalized by the addition of $N_{th}$, the PCD can almost recover its shape by squeezing, i.e., by adding $N_{S}=1$ to the Glauber-Lachs state. The squeezed photons and thermal photons act in a dramatic way such that the PCD of the SCTS  more or less resembles the PCD of a  coherent state.  To give a feel for to  what extent the PCD of resulting  SCTS resembles that of the initial coherent state, the values of widths are given in respective figures. (In fact, the recovery is much better as can be seen from Fig. 2(a) and Fig. 2(d), which is discussed below). Fig. 1(e) depicts the  PCD for  $N_{C} = 25$, $N_{S}=2$ and $N_{th} = 1$. Again, it  can be seen here, that the addition of $N_{th}=1$ to the squeezed coherent state  $N_{C} = 25$ and $N_{S}=2$ has resulted in delocalizing the the peak height of the PCD, and it is almost the same corresponding to the PCD of the SCTS for $N_{C} = 25$, $N_{S}=1$ and $N_{th} = 1$ ; however, the length of tail slightly increases. In the last Fig. 1(f), we have $N_{C} = 25$, $N_S = 2$ and $N_{th} = 2$. The peak of PCD comes down and also the tail flattens more. The addition of squeezing further  and further makes the distribution fall slightly and this was observed by Subeesh\textit{et al}.\cite{subeesh2012effect}, in the case of squeezed coherent states. Also, $N_{th}=2$ takes care of increasing the tail portion.  

This remarkable reciprocity between the squeezed photons and thermal photons is further highlighted in Fig. 2. Fig. 2(a) gives the PCD for the coherent state $N_{C} = 49$, $N_{S}=0$ and $N_{th} = 0$. In Fig. 2(b), $N_{C} = 49$,  $N_{S} = 0$ and $N_{th}=1$;  the peak of the PCD comes down to half of the initial peak, as in the Fig $1$(b). In Fig. 2(c), $N_{C} = 49$, $N_{S}=1$ and $N_{th}=0$; the peak increases by almost $100{\%}$ of its initial peak. Here, the localization of the PCD is sharply prominent, since the ratio $N_{S}/N_{C}$ is much smaller than corresponding to Fig. 1(c); yet the localization is $100{\%}$. In Fig. 2(d), $N_{C} = 49$, $N_{S} = 1$ and $N_{th} = 1$, i.e., either $N_{th}=1$ is added to the squeezed coherent state corresponding to $N_{C}=49$ and $N_{S}=1$ or $N_{S}=1$ is added to the Glauber-Lachs state for $N_{C}=49$ and $N_{th}=1$. The resulting  PCD for the SCTS (corresponding to  $N_{C} = 49$, $N_{S}=1$ and $N_{th} = 1$) is  almost like the PCD of the coherent state (corresponding to $N_{C} = 49$, $N_{S}=0$ and $N_{th} = 0$);  again, as in Fig. 1(d), the right side tail shows up. It can be seen that the difference between the widths in Fig. 2(a) and Fig. 2(d) is much smaller than the difference between the widths in Fig. 1(a) and Fig. 1(d). This means the localizing effect of squeezing on a coherent state  is  nearly compensated by the flattening of PCD by $N_{th}=1$. This also means, when the PCD of a coherent state is delocalized by the addition of $N_{th}=1$, the PCD  almost recovers its shape by squeezing, i.e., by adding $N_{S}=1$ to the Glauber-Lachs state. The squeezed photons and thermal photons act in a conspicuous way such that the PCD of a SCTS is more or less recovered and it resembles very much the PCD of a  coherent state for $N_{C}=49$. Fig. 2(e) portrays the  PCD for  $N_{C} = 49$, $N_{S}=2$ and $N_{th} = 1$. We see here, that the addition of $N_{th}=1$ to the squeezed coherent state  $N_{C} = 49$ and $N_{S}=2$ has resulted in delocalizing  the PCD and its peak height is almost the same as corresponding to the PCD of the SCTS for $N_{C} = 49$, $N_{S}=1$ and $N_{th} = 1$ ; however, the length of tail slightly increases. In the last Fig. 2(f), we have $N_{C} = 49$, $N_S = 2$ and $N_{th} = 2$. The peak of PCD comes down and also the tail flattens more. The addition of squeezing further and further makes the distributions fall slightly as observed in Fig. 2(f). Also, $N_{th}=2$ takes care of increasing the tail portion of the PCD akin to Fig. 1(f).

\section{Atomic Inversions}
To investigate the atomic inversion of the atom-field interactions, we use the famous Jaynes-Cummings model. The Jaynes-Cummings interaction Hamiltonian for the atom-field interaction is\cite{jaynes1963comparison}
\begin{equation}
\hat{H}=\hslash {\omega} \hat{a}^{\dagger}\hat{a}+\frac{\hslash\omega_{0}}{2}\hat{\sigma_z}+\hslash \lambda(\hat{\sigma}_{+}\hat{a}+\hat{\sigma}_{-}\hat{a}^{\dagger}),
\end{equation}
where $ \hat{\sigma}_{+} $ ,  $ \hat{\sigma}_{-} $  and $ \hat{\sigma}_{z} $  are the Pauli pseudospin operators; $\hat{a}$ and $\hat{a}^\dagger$ are the photon
annihilation and creation operators; $\lambda$ is the coupling constant describing the atom-field interaction; $\omega$ is the field frequency and $\omega_{0}$ is the atomic transition frequency.

\begin{figure}
\includegraphics[width = \linewidth]{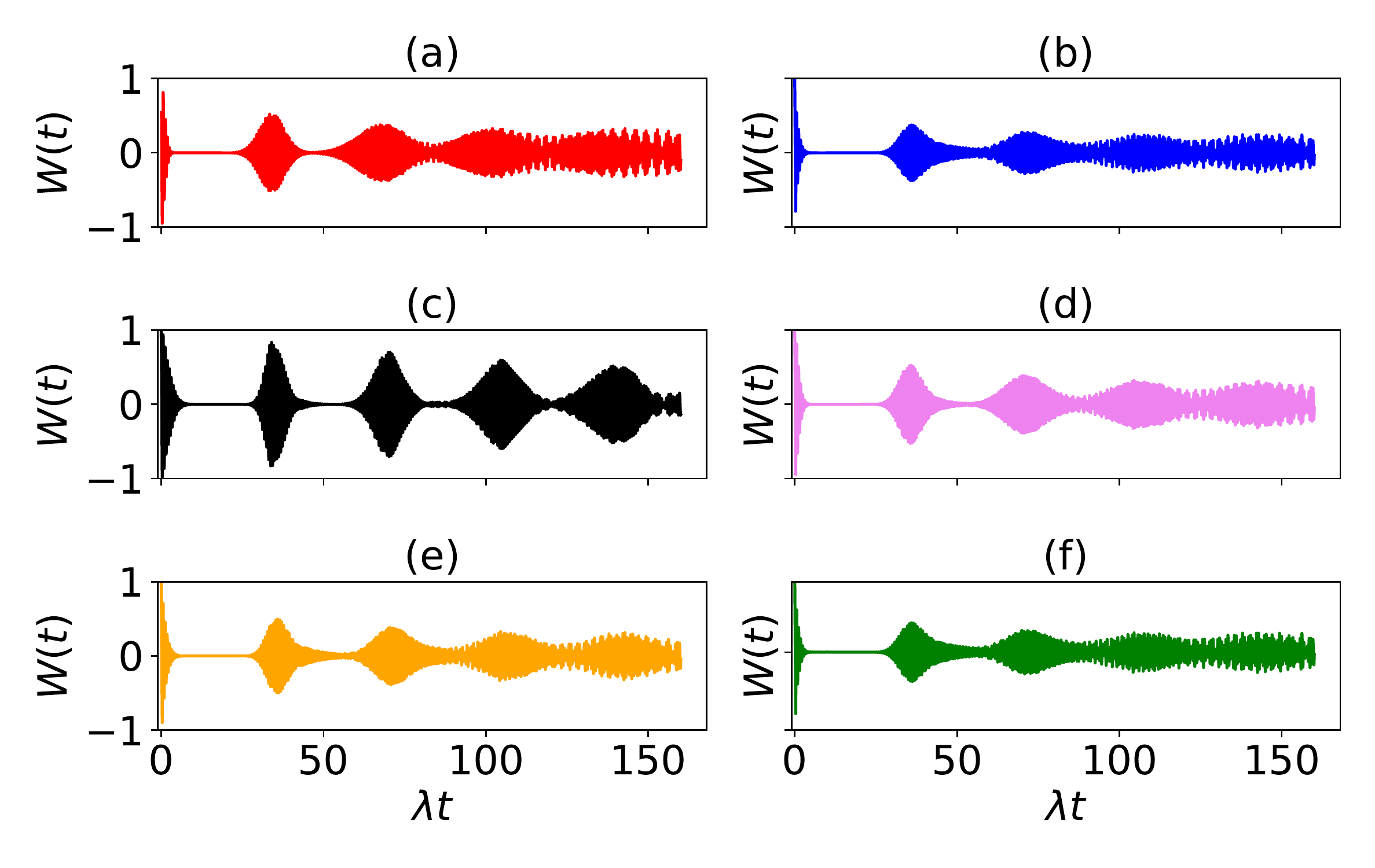}
\caption{tomic inversion for SCTS for $(N_{C}, N_{S}, N_{th})$ = (a)$(25, 0, 0)$, (b)$(25, 0, 1)$, (c) $(25, 1, 0)$,
(d) $(25, 1, 1)$, (e) $(25, 2, 1)$, (f) $(25, 2, 2)$.}
\label{fig3}
\end{figure}
	
\begin{figure}
\includegraphics[width = \linewidth]{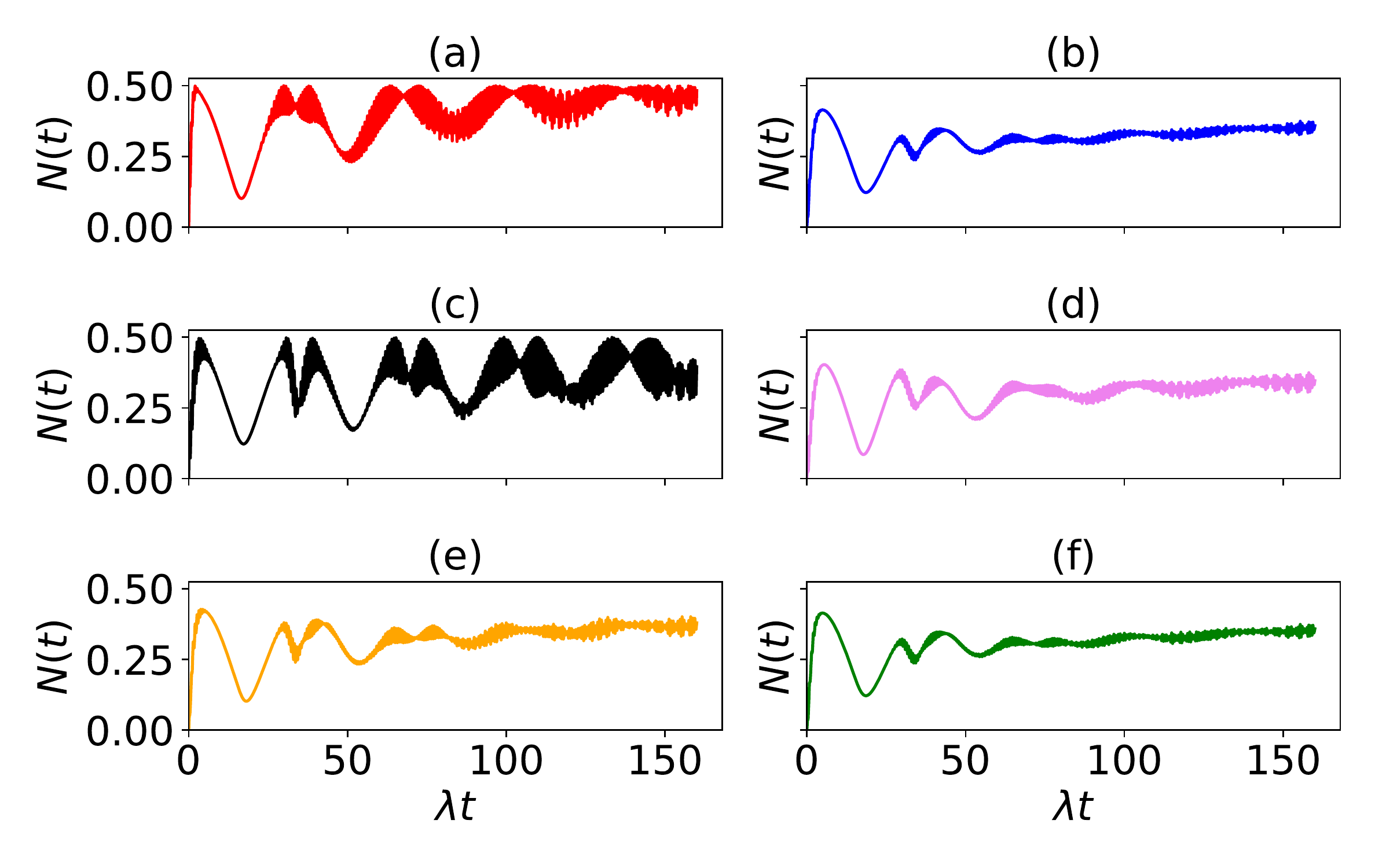}
\caption{Atomic inversion for SCTS for $(N_{C}, N_{S}, N_{th})$ = (a)$(49, 0, 0)$, (b)$(49, 0, 1)$, (c) $(49, 1, 0)$,
(d) $(49, 1, 1)$, (e) $(49, 2, 1)$, (f) $(49, 2, 2)$.}
\label{fig4}
\end{figure}

The quantity of interest which describes the dynamics of interaction between the radiation field and atom is the atomic inversion, which is defined as the difference in the probabilities of finding the atom in the excited state and ground state and it is given by\cite{gerry2005introductory}

\begin{eqnarray}
 W(t)= \sum_{n=0}^{\infty}P(n)\cos(2\lambda\sqrt{n+1}~t),
\end{eqnarray}
where $P(n)$ is the PCD of the initial state of the radiation field and the atom is initially supposed to be in the excited state.

Consequences of such dramatic tussle between the thermal photons and squeezed photons discussed in Section 2 above are presented in the plots for the atomic inversion  $W(t)$ in  Figs. 3 and 4 for different combinations of $N_{C}$, $N_{S}$ and $N_{th}$ corresponding to Figs. 1 and 2.  Fig. 3(a) represents the atomic dynamics of a coherent state for $N_{C}=25$, viz., the familiar collapse and revival phenomena associated with a two-level atom interacting with a coherent state, whose PCD is given in Fig. 1(a). Fig. 3(b) gives the plot of $W(t)$ corresponding to the radiation field for which $N_{C} = 25, N_{S} = 0, N_{th} = 1$, whose PCD is given in Fig. 1(b), i.e., one mean thermal photon is added to the coherent state of $N_{C}=25$. Here, we observe that the amplitude of revival part of the dynamics decreases and the second collapse time of the dynamics disappears. This dynamics of atom-field interactions was studied by Satyanarayna \textit{et al.}\cite{satyanarayana1992glauber}. Fig. 3(c) gives the atomic dynamics for the squeezed coherent state for which $N_{C} = 25, N_{S} = 1, N_{th} = 0$, whose PCD is in Fig. 1(c). The atomic dynamics for such squeezed coherent states is well studied in \cite{subeesh2012effect}.  Fig. 3(d) gives the plot of $W(t) $ for the SCTS for which $N_{C} = 25, N_{S} = 1, N_{th} = 1$, whose PCD is in Fig. 1(d). This is the state in which the addition of squeezed photons compensates the broadening of the Glauber-Lachs state or equivalently the sharply localized squeezed coherent state is delocalized again by the addition of one mean thermal photon. It is observed that squeezing corresponding to $N_{S}=1$ virtually cancels the effects of addition of thermal photons $N_{th}=1$ at the level of distributions and associated atomic inversions. Mutually compensatory struggles between the thermal photons and squeezed photons are well reflected in - Figs. $1$(a) to $1$(d) at the level of distributions - and, Figs. $3$(a) to $3$(d) at the level of atomic inversions. It is remarkable that the addition of one mean thermal and  one mean squeezed photon to the coherent state does not change much either the distributions or the atomic dynamics and this happens via the localization and delocalization effects on the distributions due to squeezed photons and thermal photons, respectively. On increasing the squeezing (for $N_{C} = 25$, $N_{S}=2$ and $N_{th} = 1$ ), the atomic dynamics tends towards that of a squeezed coherent state, as in  Fig. 3(e). On increasing the mean thermal photons (for $N_{C} = 25, N_{S} = 2, N_{th} = 2$), the atomic dynamics becomes more thermal as in Fig. 3(f). This is evident from the shortening of collapse times. 

For $N_S = 1$ and $N_{th} = 1$, it is observed that  the dynamics is similar to that of  a coherent state; but in the case of $N_S = 2, N_{th} = 2$, it is seen that thermal photons dominate over squeezed photons. This shows (see Fig. 3(f)) that addition of equal number of mean thermal and squeezed photons does not always result in the dynamics similar to coherent state dynamics; actually, the dynamics is more akin to  the domination by thermal photons.

To highlight the balancing between squeezing and thermal photons, the atomic dynamics is depicted in  Fig. 4 for $N_{C} = 49$, for the same combinations of $N_{S}$ and $N_{th}$ corresponding to the distributions in Fig. 2. It is observed that the atomic dynamics for the SCTS state corresponding to  $N_{C} = 49$, $N_{S}=1, N_{th}=1$ is practically same as the atomic dynamics for the coherent state corresponding to $N_{C}=49$. As in the earlier case, the addition of an equal number of mean thermal and squeezed photons results in a dynamics, in which, the effects of thermal photons dominate, as shown in Fig. 4(f). So, in the tussle between the thermal photons and squeezed photons, it seems that, the effects of thermal photons have an upper hand both at the level of PCD and $W(t)$.  

\section{Entanglement Dynamics}

\begin{figure}
\includegraphics[width = \linewidth]{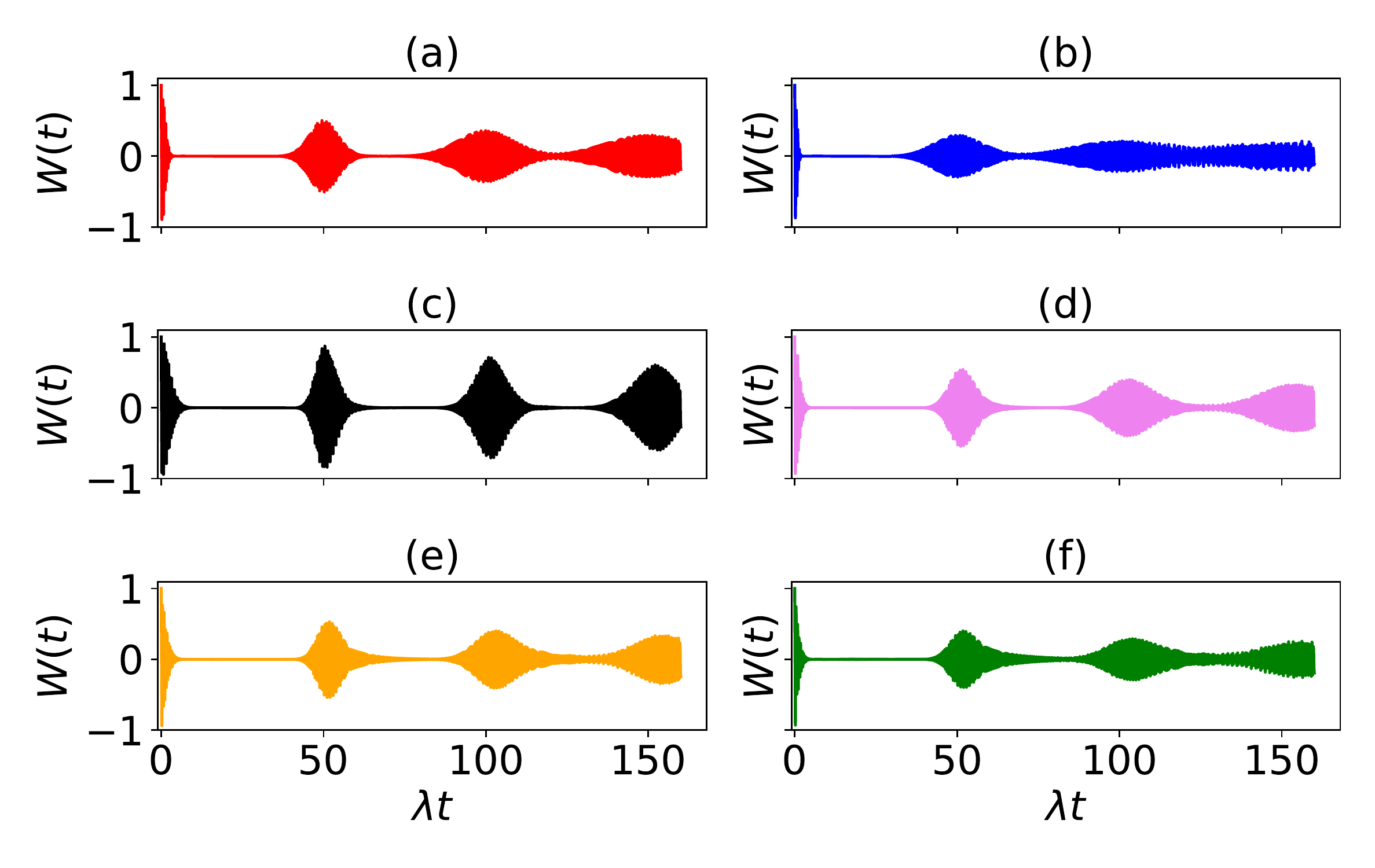}
\caption{Entanglement dynamics for SCTS for $(N_{C}, N_{S}, N_{th})$ = (a) $(25, 0, 0)$, (b) $(25, 0, 1)$, (c) $(25, 1, 0)$, (d) $(25, 1, 1)$, (e) $(25, 2, 1)$, (f) $(25, 2, 2)$.}
\label{fig5}
\end{figure}
		
\begin{figure}
\includegraphics[width = \linewidth]{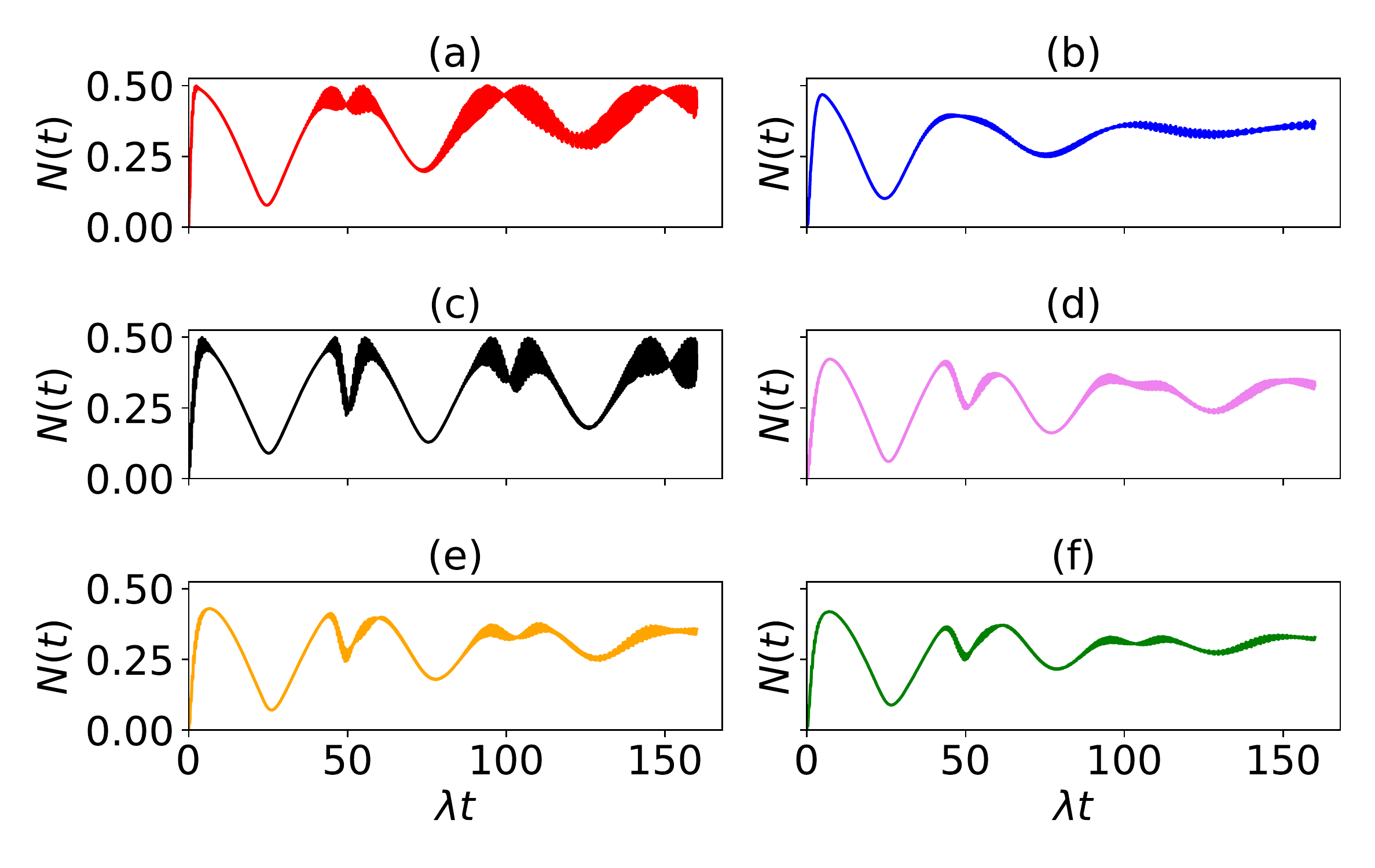}
\caption{Entanglement dynamics for SCTS for $(N_{C}, N_{S}, N_{th})$ = (a) $(49, 0, 0)$, (b) $(49, 0, 1)$, (c) $(49, 1, 0)$, (d) $(49, 1, 1)$, (e) $(49, 2, 1)$, (f) $(49, 2, 2)$.}
\label{fig6}
\end{figure}
			
The next quantity, that we are interested in is to investigate the entanglement dynamics of the atom-field interactions. There are various quantities to measure the entanglement between two systems, like, von Neumann entropy and linear entropy, etc.
Here, to measure the entanglement between the atom and the field, negativity $N(t)$ has been used, which is defined as the absolute sum of the negative eigenvalues of the partially transposed density operator $\hat{\rho}_{\text{tot}}^{\text{PT}}$ \cite{wei2003maximal}.
If, $\lambda_{k}$ are the eigenvalues of $\hat{\rho}_{\text{tot}}^{\text{PT}}$, then $N(t)$ is given by
\begin{equation}
N(t)=\sum_{k}\left[|\lambda_{k}|-\lambda_{k}\right]/2.
\end{equation}
Figs. 5 and 6 depict the entanglement dynamics for SCTS. Here, Fig. 5(a) represents the plot of $N(t)$ for the coherent state $N_{C} = 25$, $N_{S} = 0$ and $N_{th} =0$. Initially, the radiation field and the atom are unentangled. However, $N(t)$ describes the maximum entanglement between the radiation field and the atom at various instants of time. Fig. 5(b) corresponds to $N_{C} = 25$, $N_{S} = 0$ and $N_{th} =1$, for which the dynamics of $N(t)$ changes very significantly and the amplitude reduces. The addition of thermal photons very markedly reduces the field-atom entanglement. In other words, thermal photons `wash off' the entanglement. On contrary,  if only the squeezed photons are added, as in Fig. 5(c), the dynamics is similar to the $N(t)$ of the squeezed coherent state, which was studied by Subeesh \textit{et al}.\cite{subeesh2012effect}. Now, for $N_{C} = 25$, $N_{S} = 1$ and $N_{th} =1$, the dynamics is very interesting. In the case of atomic inversion, for the same combination of  $N_{C}$, $N_{S}$, $N_{th}$, we observe that the dynamics of $W(t)$ resembles very much like that of a  coherent state. For the entanglement dynamics that is not the case - the $N(t)$ in Fig. 5(d) is more akin to that corresponding to a Glauber-Lachs state as in Fig. 5(b). So, the addition of thermal photons has contrasting effects on $W(t)$ and $N(t)$. On increasing the squeezing  further, the entanglement dynamics in Fig. 5(e) retains the characteristics of Fig. 5(d), which means the aspects of $N(t)$ which were washed off by thermal photons could not be recovered. Fig. 5(f) corresponds to $N_{C} = 25$, $N_{S} = 2$ and $N_{th} =2$ and the entanglement dynamics is more like that of a Glauber-Lachs state. 

The entanglement dynamics pattern for $N_{C} = 49$ has similar behaviour  to those for  $N_{C} = 25$, as can be seen by comparing Fig. 5 with Fig. 6.

\section{Atomic inversion and entanglement dynamics for coherent squeezed thermal states}

\subsection{Photon counting distribution}
Having studied the atomic inversion and entanglement dynamics for squeezed coherent thermal state (SCTS), it is appropriate to study the dynamics for coherent squeezed thermal states(CSTS). The density operator for CSTS is given by\cite{PhysRevA.47.4474}

\begin{equation}
\hat{\rho}_{\text{CST}} = \hat{S}(\zeta)\hat{D}(\alpha)\hat{\rho}_{th}\hat{D}^{\dagger}(\alpha)\hat{S}^{\dagger}(\zeta).
\end{equation}
Here, the order of the displacement and squeezing operators as in the definition of SCTS are exchanged. $\hat{S}(\zeta)$ and $\hat{D}(\alpha)$ are related as following
\begin{equation}
    \hat{S}(\zeta) \hat{D}(\alpha) = \hat{D}(\alpha \cosh r - \alpha^*\exp (i \varphi )\sinh r) \hat{S}(\zeta).
\end{equation}

Fig. 7 shows the plots of the PCD for CSTS. Fig. 7(a) gives the distribution for the coherent state $N_{C} = 25$, $N_{S}=0$, $N_{th} =0$. On the addition of one mean thermal photon to this coherent state, the resulting distribution is given in Fig. 7(b). As expected the height of PCD falls by half and also broadens. However, for $N_{C} = 25$, $N_{S}=1$, $N_{th} =0$, i.e., on  adding one mean squeezed photon, the PCD in Fig. 7(c) gets localized very remarkably, as the peak increases by four times. This is very interesting because in the case of SCTS, the enhancement of peak height was only double. Although, the two operators $\hat{S}(\zeta)$ and $\hat{D}(\alpha)$ are connected only by a shifting parameter (see Eq.(24)), these states show very sharp localisation of the PCDs, when mean squeezed photons are increased. On the other hand, if mean thermal photons are increased, the effect is same as that for SCTS, viz., to reduce the height of PCD, and at the same time broadening it. In general, addition of squeezed photons make the peaks of  PCD plots far taller than that for SCTS.

\begin{figure}
\centering
\includegraphics[width = \linewidth]{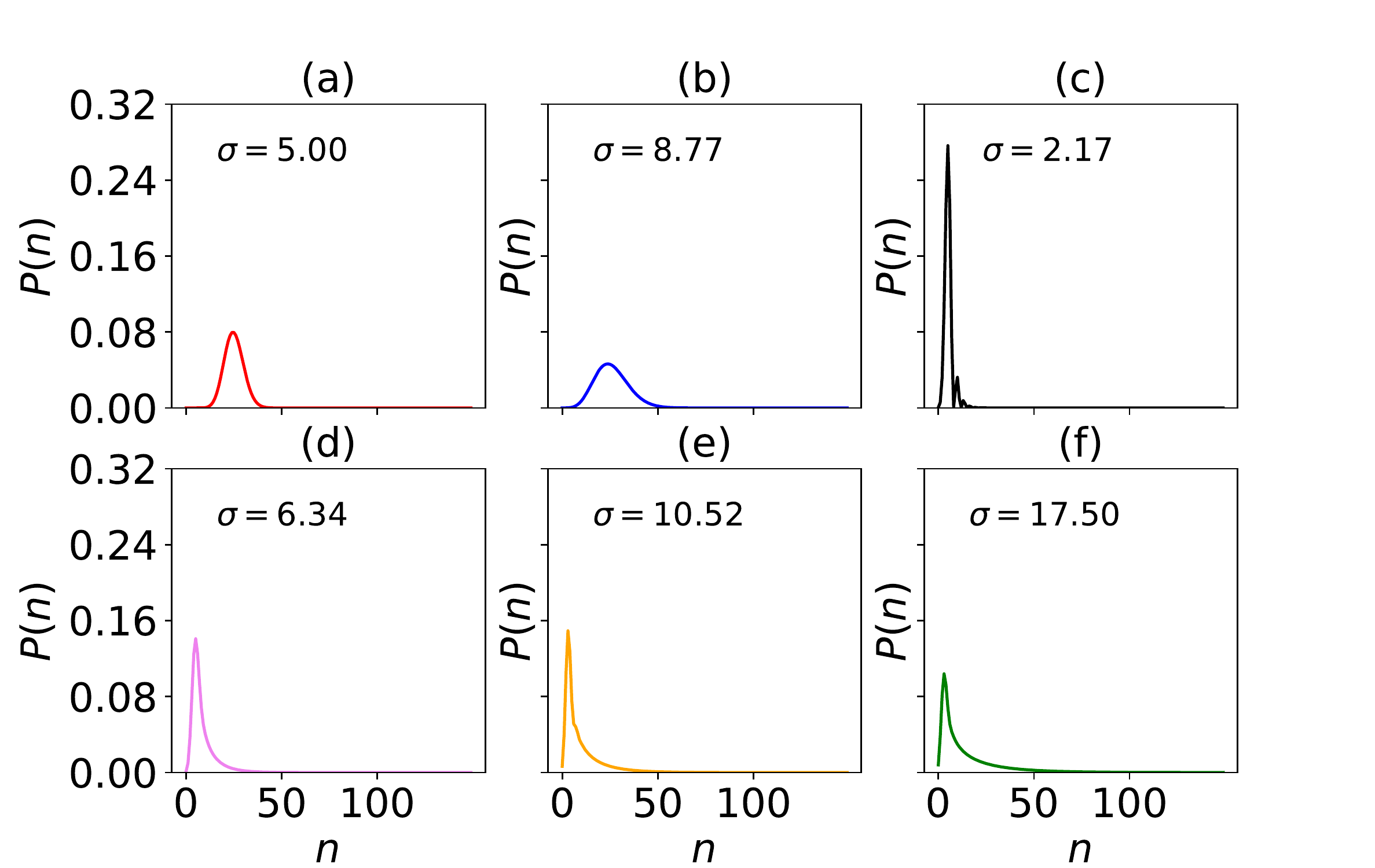}
\caption{Photon counting distributions for CSTS for $(N_{C}, N_{S}, N_{th})$ = (a) $(25, 0, 0)$, (b) $(25, 0, 1)$, (c) $(25, 1, 0)$, (d) $(25, 1, 1)$, (e) $(25, 2, 1)$, (f) $(25, 2, 2)$.}
\label{fig7}
\end{figure}

\subsection{Atomic inversion and entanglement dynamics}

\begin{figure}
\centering
\includegraphics[width = \linewidth]{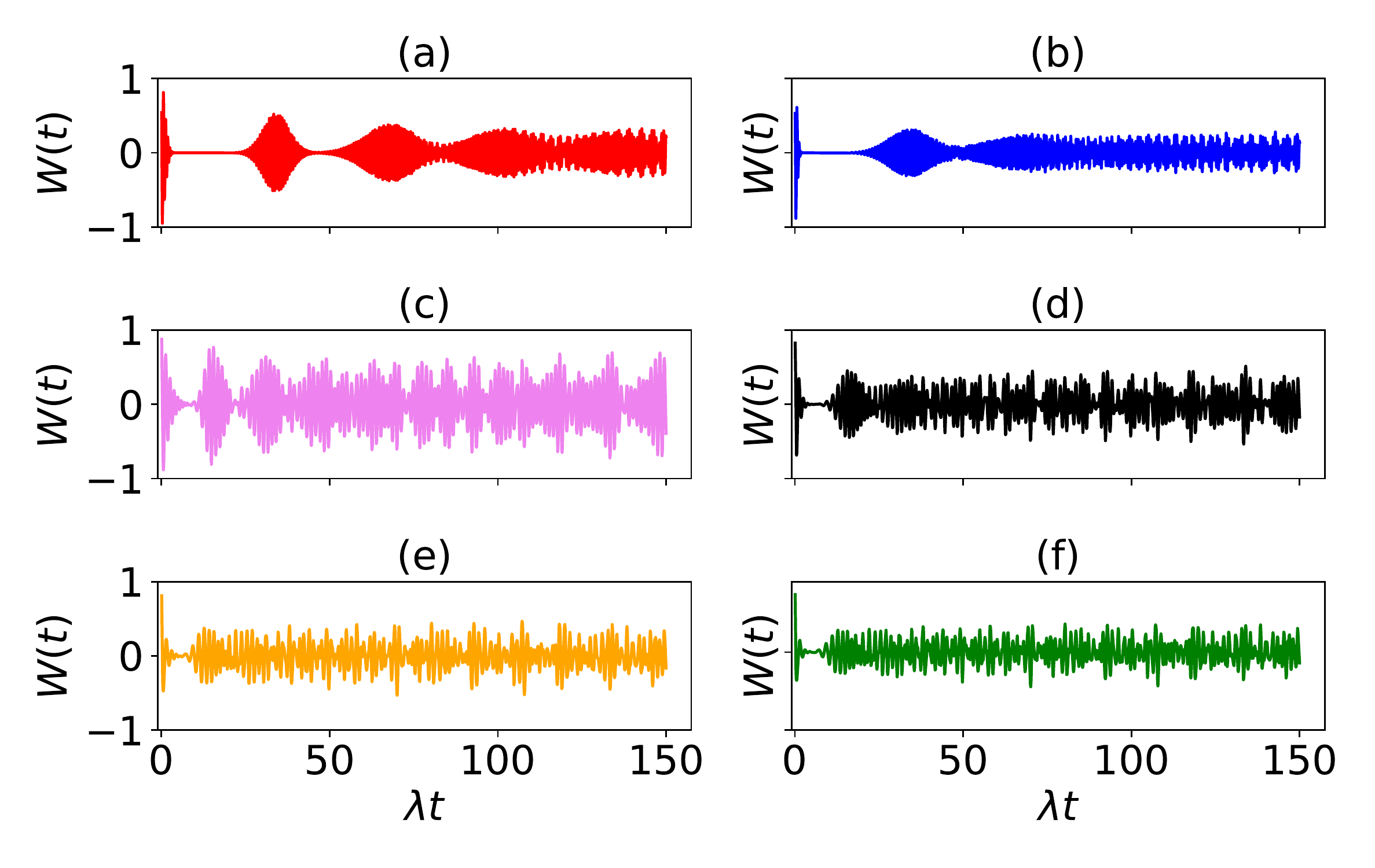}
\caption{Atomic inversion for CSTS for  $(N_{C}, N_{S}, N_{th})$ = (a) $(25, 0, 0)$, (b) $(25, 0, 1)$, (c) $(25, 1, 0)$, (d) $(25, 1, 1)$, (e) $(25, 2, 1)$, (f) $(25, 2, 2)$.}
\label{fig8}
\end{figure}

\begin{figure}
\centering
\includegraphics[width = \linewidth]{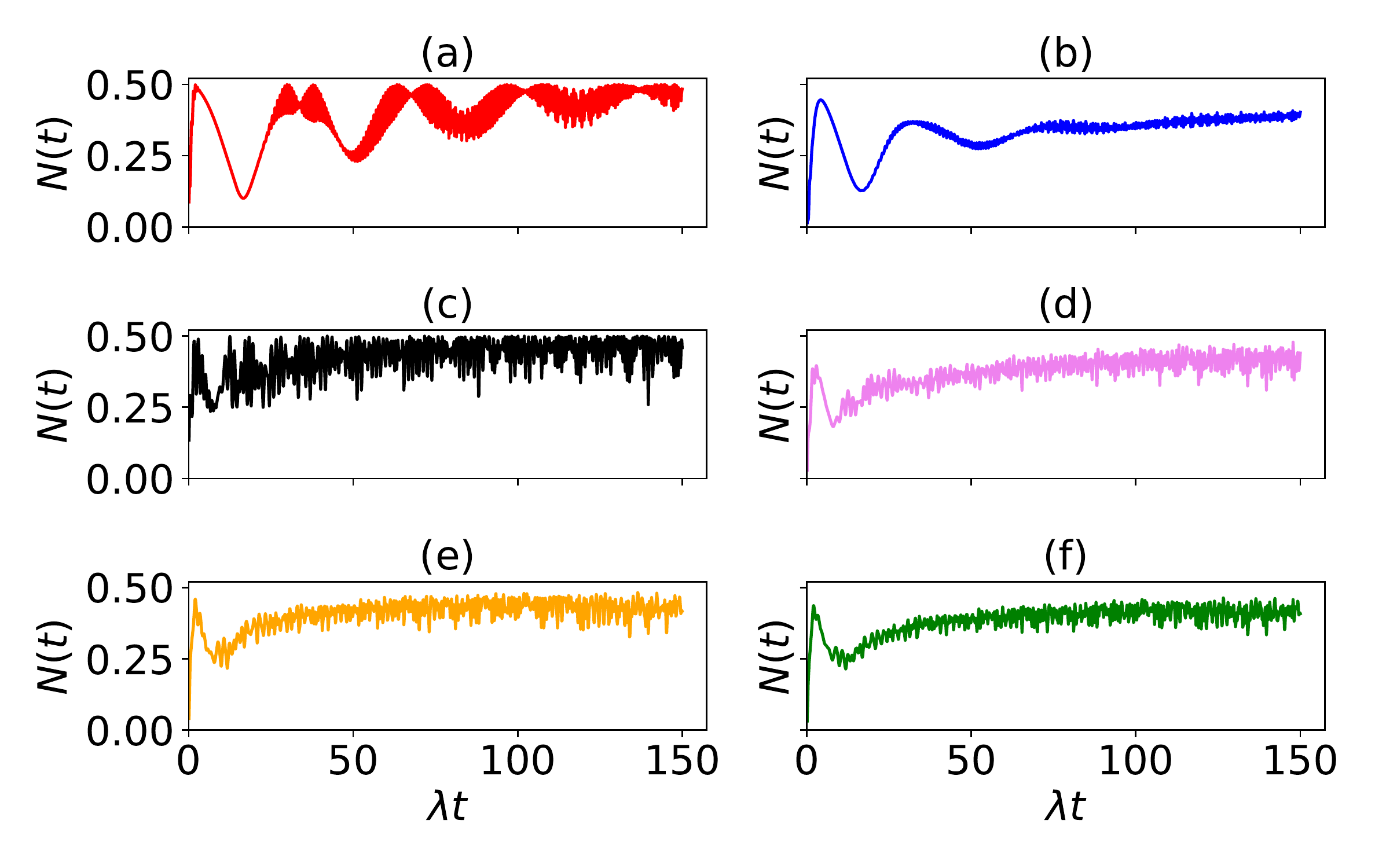}
\caption{Entanglement dynamics for CSTS for  $(N_{C}, N_{S}, N_{th})$ = (a) $(25, 0, 0)$, (b) $(25, 0, 1)$, (c) $(25, 1, 0)$, (d) $(25, 1, 1)$, (e) $(25, 2, 1)$, (f) $(25, 2, 2)$.}
\label{fig9}
\end{figure}

Fig. 8 represents the atomic inversion for CSTS. For this state, the addition of mean squeezed photons has very different effect on the dynamics as compared to the dynamics for SCTS . In this case also, $W(t)$ starts from the dynamics of a coherent state. Then, if one mean thermal photon is added to the field, we get almost the same pattern as we have observed for SCTS (see Fig. 8(b)). But, when we add a single mean  squeezed photon to the field, very different pattern is observed as in Fig. 8(c). The first collapse duration becomes very short, though the first revival is sharp.  The subsequent collapses disappear.  Fig. 8(d) corresponds to the CSTS $ (N_{C}=25, N_{S}=1, N_{th}=1)$, for which $W(t)$ is  noisy. It is to be contrasted with the $W(t)$ of SCTS corresponding to the same values $(N_{C}=25, N_{S}=1, N_{th}=1)$, for which the collapses and revivals very much remain, resembling the $W(t)$ of a coherent state (see Fig. 3(d)). It seems, for CSTS, the mean thermal and squeezed photons collectively interplay to destroy the collapse-revival pattern corresponding to a coherent state. On further increasing the number of  mean thermal or squeezed photons to the field, $W(t)$ (in Figs. 8(e) and Fig. 8(f)) show the almost same noisy pattern with decreased amplitude as in Fig. 8(d).

The entanglement dynamics $N(t)$ for CSTS is plotted in Fig. 9. Like atomic inversion, $N(t)$ also shows different behaviour for CSTS compared to SCTS. Here also, $N(t)$ starts from the dynamics of a coherent state interacting with the two-level atom which is plotted in Fig. 9(a). It shows almost similar pattern as observed in the case of SCTS, when we add one mean thermal photon to the field ($N_{C}=25, N_{S}=0, N_{th}=1$)(see Fig. 9(b)). The addition of a mean thermal photon reduces the field-atom entanglement very significantly. Here again, the thermal photons `wash off' the entanglement.  On the addition of a mean squeezed photon to the field, i.e., ($N_{C} = 25, N_{S}=1, N_{th}=0$), $N(t)$ shows very noisy behaviour (Fig. 9(c)). The collapse-revival pattern of the dynamics looks drastically different - the collapses and revivals disappear.  Next, when one mean thermal photon and one mean squeezed photon are successively added to the field which corresponds to ($N_{C} = 25, N_{S}=1, N_{th}=1$), $N(t)$ in Fig. 9(d) shows a noisy dynamics-as if, the patterns in Fig. 9(b) and Fig. 9(c) are superposed. This noisy $N(t)$ of CSTS is to be contrasted with more thermal dominated $N(t)$ of SCTS as in Fig. 5(d). On further addition of  mean squeezed photons or mean thermal photons, the dynamics of $N(t)$ does not change much as evident in Fig. 9(e) and 9(f).

From the above  dynamics, it is to be observed that, if mean thermal photons are added to the coherent field, $W(t)$ and $N(t)$ show almost similar behaviour for the fields SCTS and CSTS respectively. But, they behave very differently, if we add mean squeezed photons to the coherent field. So, the atomic dynamics $W(t)$ and the entanglement dynamics $N(t)$ for these states are more sensitive to the addition of squeezed photons as compared to the addition of thermal photons.

\section{Second order correlation function $G^{2}(0)$ for SCTS and CSTS}

The second order correlation function $G^{2}(0)$ is given by\cite{PhysRevA.47.4474}
\begin{equation}
G^{2}(0) = \frac{\langle n^{2} \rangle - \langle n \rangle}{{\langle n \rangle}^{2}}.
\end{equation}

\begin{figure}
\centering
\includegraphics[width = \linewidth]{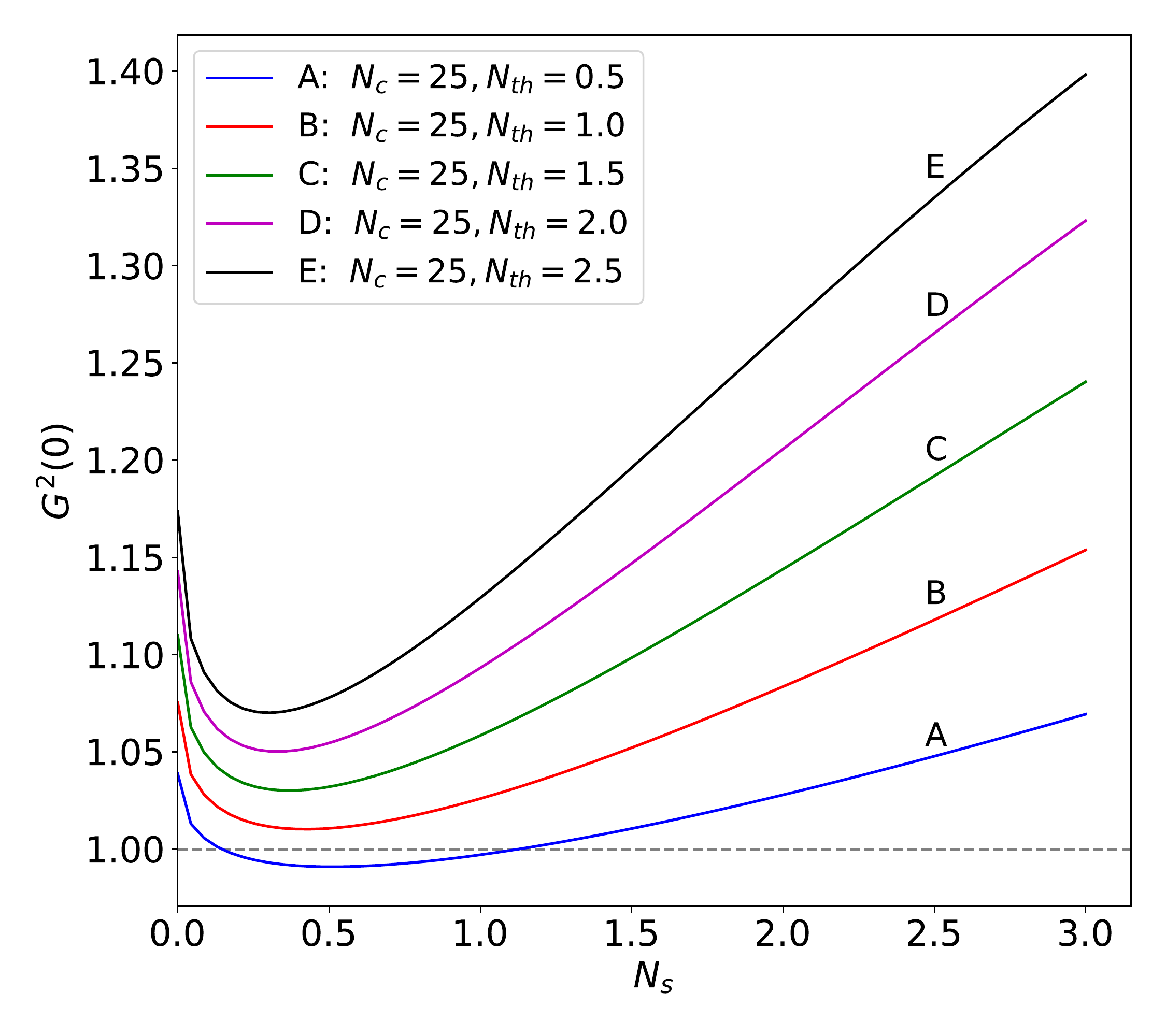} 
\caption{$G^{2}(0)$  $vs$ $N_{S}$ for SCTS.}
\label{fig10}
\end{figure}

\begin{figure}
    \centering
    \includegraphics[width = \linewidth]{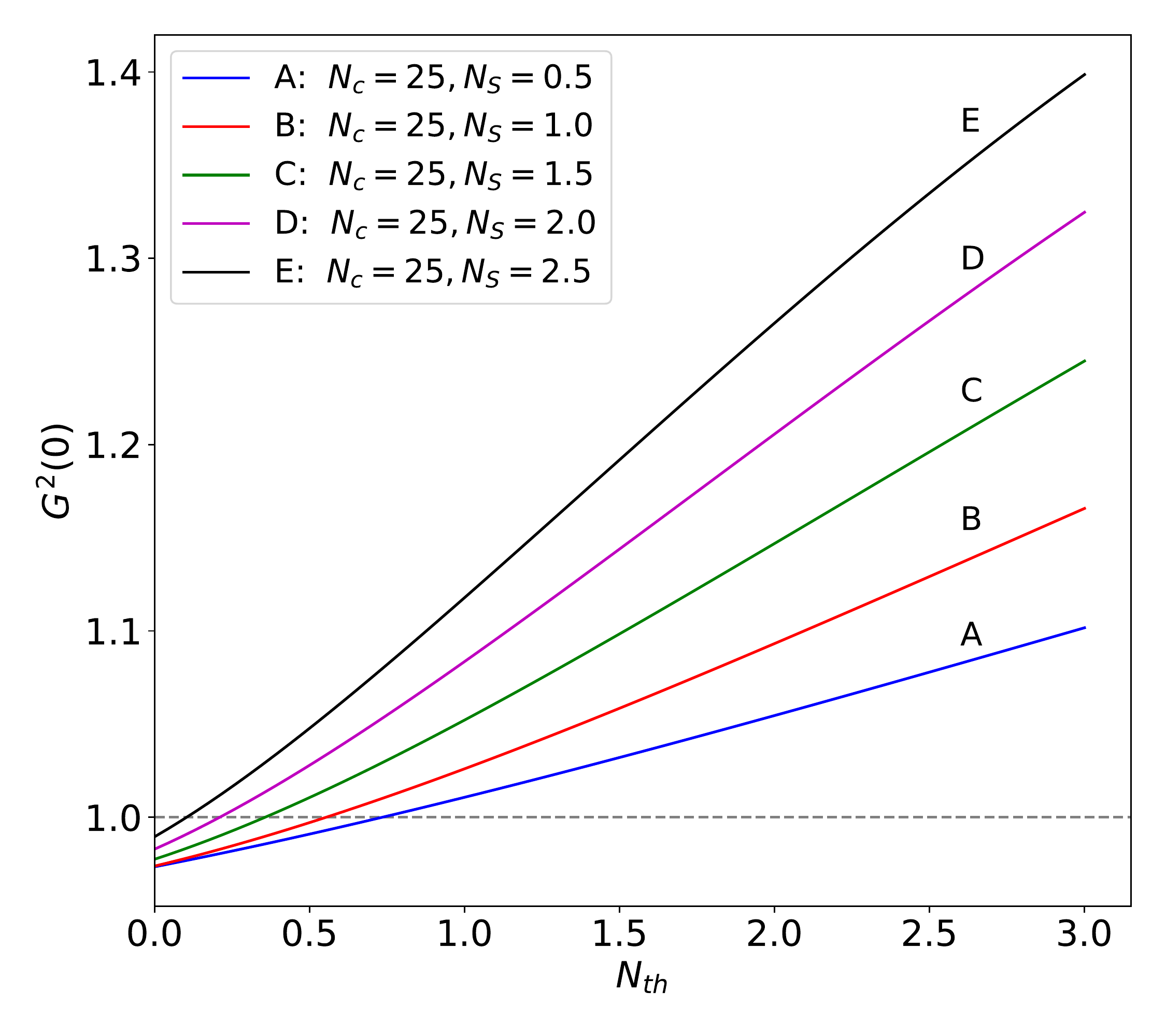}
    \caption{$G^{2}(0)$  $vs$ $N_{th}$ for SCTS.}
    \label{fig:my_label}
    \label{fig11}
\end{figure}

\begin{figure}
\centering
\includegraphics[width = \linewidth]{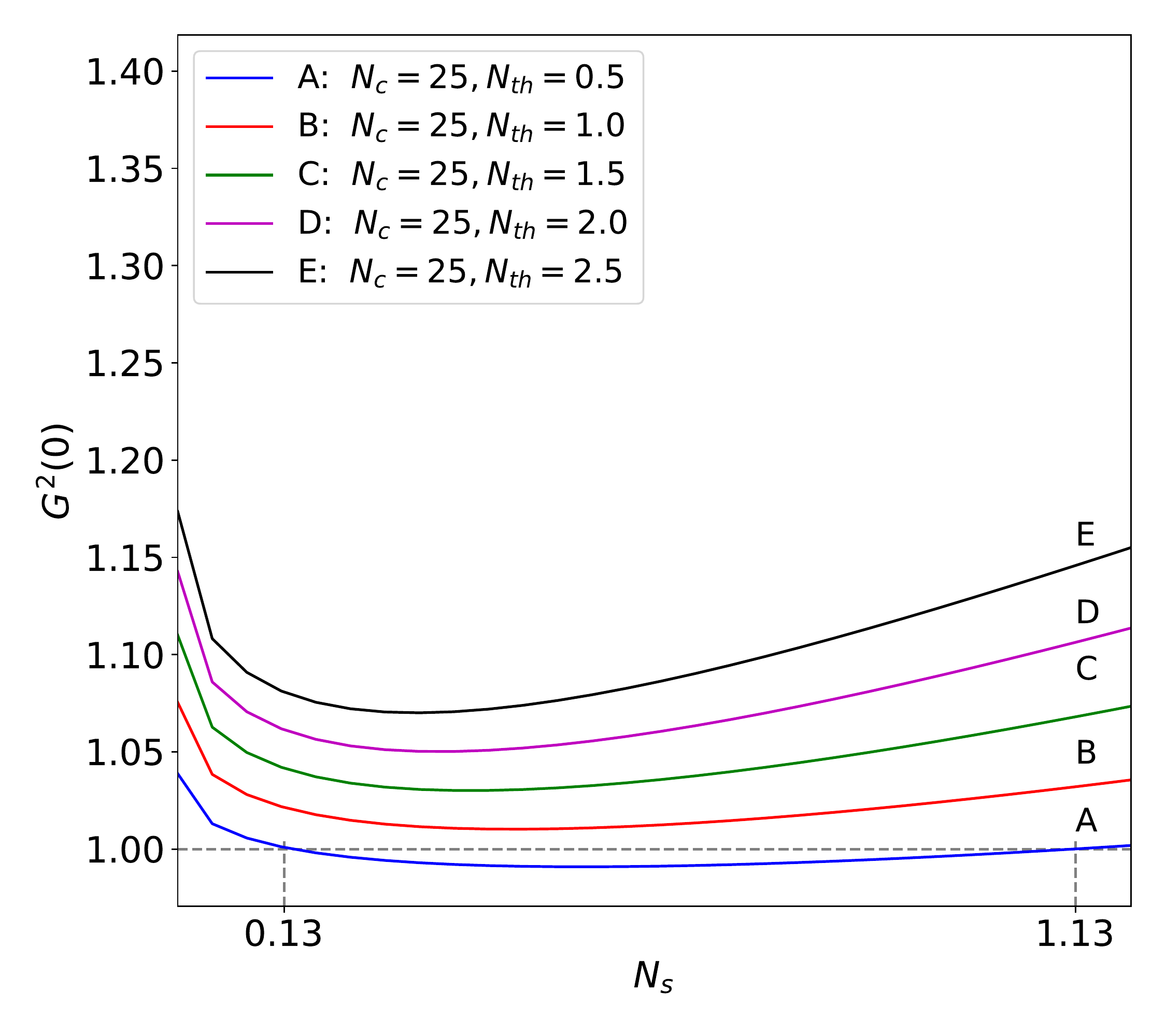}
\caption{Zoomed portions of $G^{2}(0)$  $vs$ $N_{S}$ for SCTS.}
\label{fig12}
\end{figure}

\begin{figure}
    \centering
    \includegraphics[width = \linewidth]{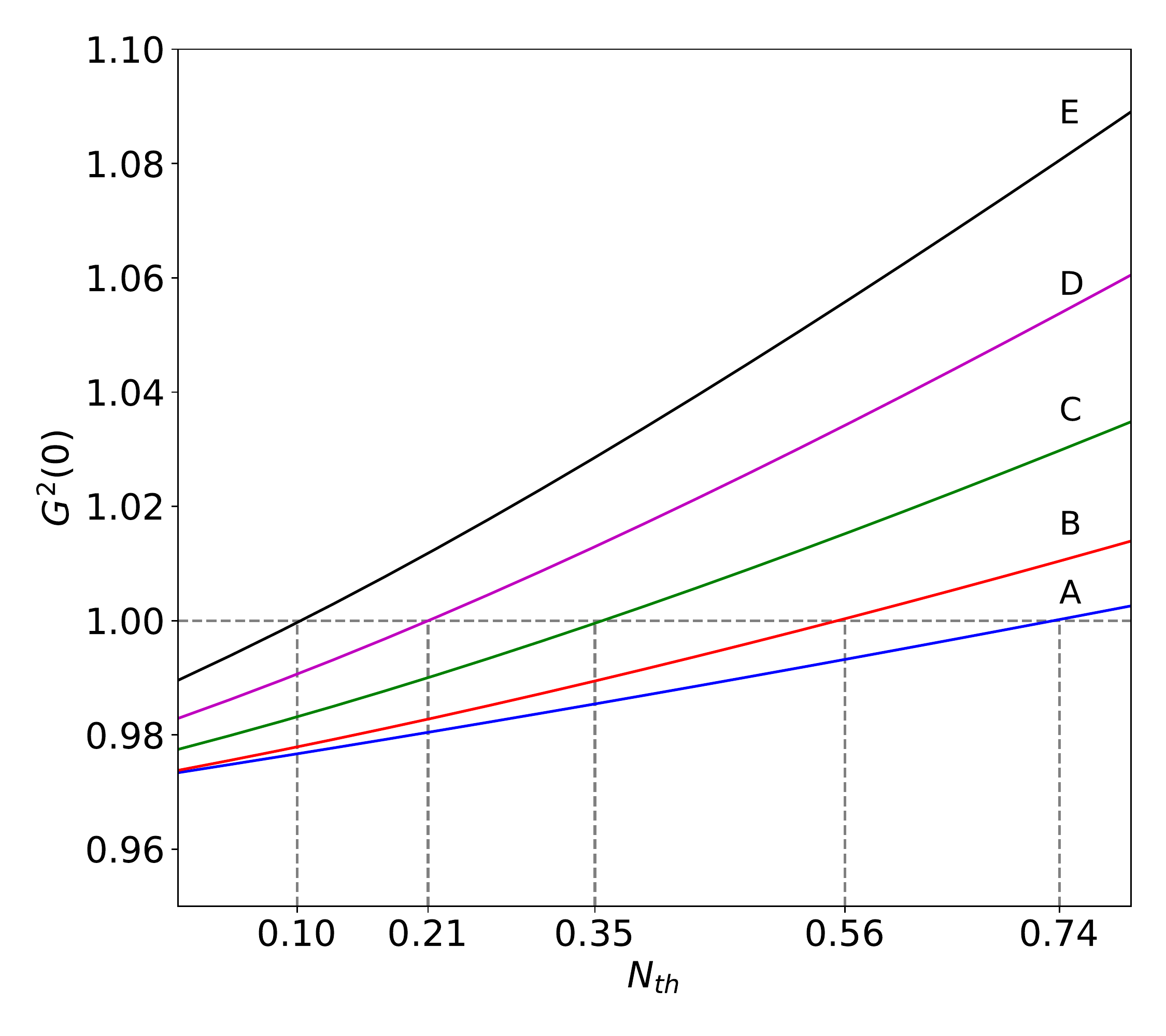}
    \caption{Zoomed portions of  $G^{2}(0)$  $vs$ $N_{th}$ for SCTS.}
    \label{fig13}
\end{figure}

\begin{figure}
\centering
\includegraphics[width = \linewidth]{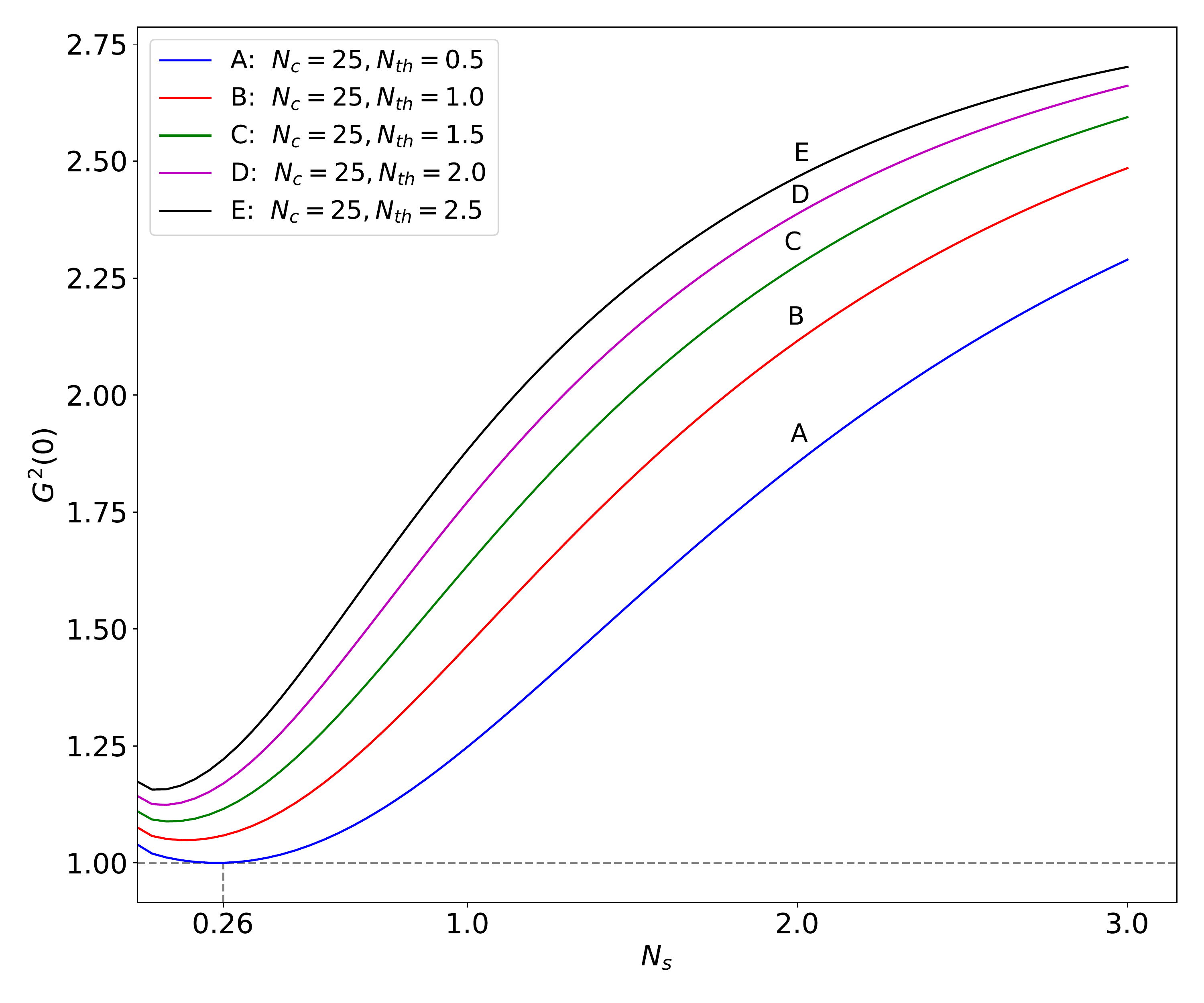}
\caption{$G^{2}(0)$  $vs$ $N_{S}$ for CSTS.}
\label{fig14}
\end{figure}

\begin{figure}
    \centering
    \includegraphics[width = \linewidth]{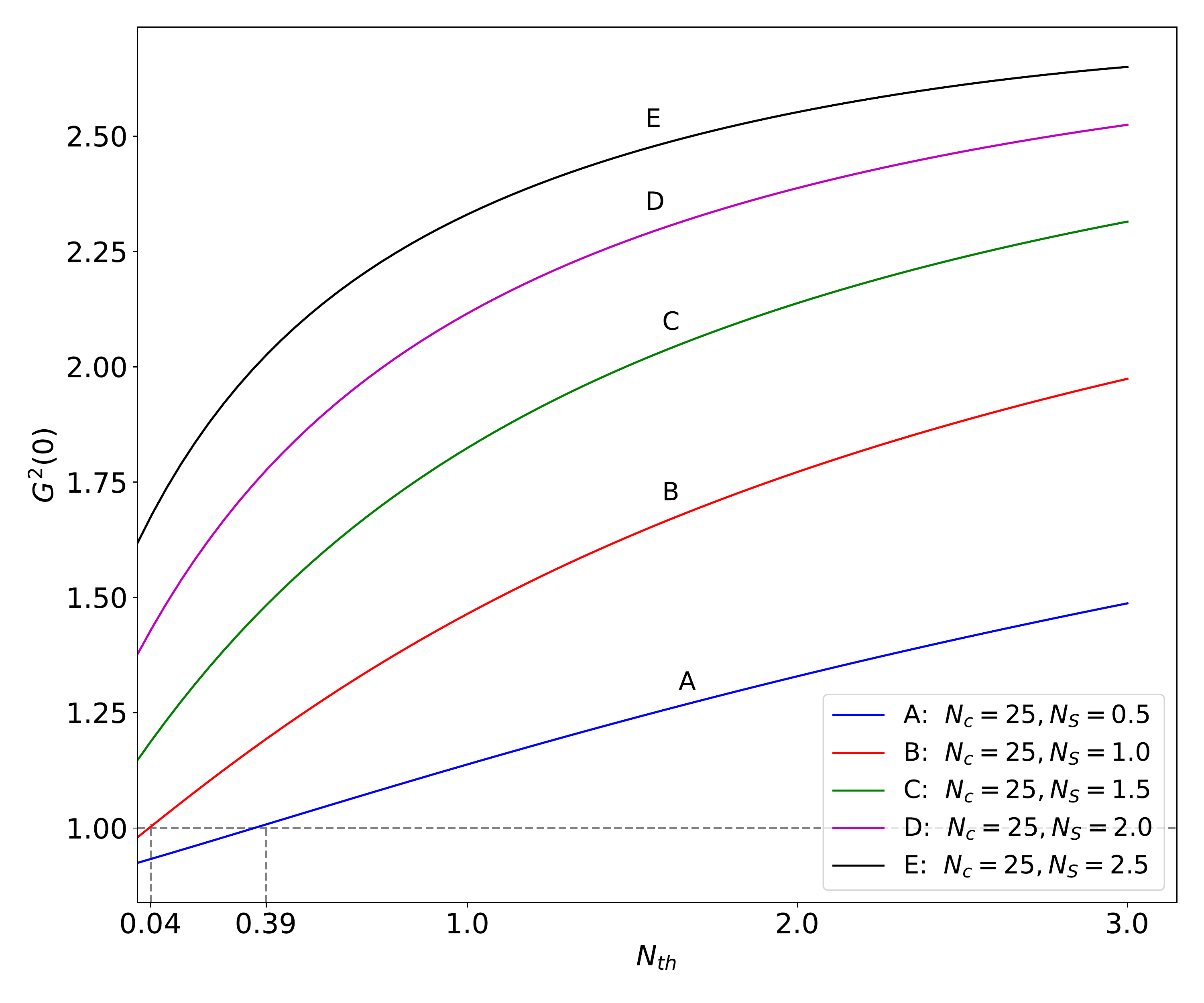}
    \caption{$G^{2}(0)$  $vs$ $N_{th}$ for CSTS.}
    \label{fig15}
\end{figure}

Figs. 10 and 11 represent the  plots of $G^{2}(0)$ $vs$ $N_{S}$ and $G^{2}(0)$  $vs$ $N_{th}$ respectively for various SCTS. In Fig. 10,  $N_C = 25$, and, then for various fixed number of $N_{th}$, $N_{S}$ is  varied. In Fig. 11, $N_C = 25$, and, then for various fixed number of  $N_{S}$,  $N_{th}$ is  varied. 

In Fig. 10, the graph A corresponding to $N_{th}=0.5$ is interesting. It starts from a value greater than $1$, as it should, because the state of the field is in a Glauber-Lachs state. On increasing $N_{S}$, $G^{2}(0) = 1$, for $N_{S} = 0.13$ and $1.13$, and such states of the radiation field exhibit zero Hanbury Brown and Twiss (HBT) correlation. For all values of $N_{S}\in [0.13,1.13]$, (see the zoomed portion in Fig. 12), $G^{2}(0) < 1$,  which indicates that the corresponding states are anti-bunched and such states are nonclassical. This behaviour of $G^{2}(0)$ can be associated with the localisation of the PCD of the SCTS.  The localisation is prominent for certain  combination of values of $N_{S}$ and $N_{th}$ (and for such combinations, $N_{S}$ dominates over $N_{th}$). For $N_{S}\in[0,0.13]$  and for $N_{S} > 1.13$, $G^{2}(0)> 1$, and the corresponding states are bunched. Clearly, the interplay between $N_{th}$ and $N_{S}$ is such that for certain combinations of $N_{th}$ and $N_{S}$, the radiation field admits antibunched states  . For other values of $N_{th}$, the graphs B to E, start from a value greater than $1$ and for a small range of values of $N_S$, $G^{2}(0)$ decreases; however, the states are bunched only. 

Fig. 11 shows the plots of  $G^{2}(0)$ $vs$ $N_{th}$. Here, $N_{C}=25$ and for various fixed values of $N_{S}$, $N_{th}$ is varied and the $G^{2}(0)$ curves are obtained as shown in graphs A-E. These graphs  indicate $G^{2}(0) < 1$, in various intervals of $N_{th}$, as can be seen in Fig. 13. It is interesting to note that, $G^2(0)= 1$ for  $N_{th} = 0.74, 0.56, 0.35, 0.21, 0.10$, respectively. So, at these values of $N_{th}$, the radiation field, shows zero HBT correlation. On increasing $N_{th}$ beyond these values, the interplay between $N_{th}$ and $N_{S}$ is complex, and the corresponding states are bunched. So, from Figs. 10 and 11, it is observed that initially for lower values of $N_{th}$ and relatively higher values of $N_{S}$, the states of radiation field are antibunched. As, $N_{th}$ is increased, the resulting states are bunched.

Figs. 14 and 15 represent the plots of  $G^{2}(0)$ $vs$ $N_{S}$ and $G^{2}(0)$ $vs$ $N_{th}$ respectively for CSTS. Again, the same values of ($N_{C}, N_{S}, N_{th}$) as in Figs. 10 and 11 have been employed. The graph A in Fig. 14 represents $G^{2}(0)$ for $N_{th} = 0.5$, and $N_{S}$ takes values in $[0,3.0]$. The radiation field possesses a zero HBT state for $N_{S}=0.26$; and, all states corresponding to other values of $N_{S}$ are bunched. For other values of $N_{th}$, $G^{2}(0) > 1$, as can be seen in  graphs B-E.

The graph A in Fig. 15, represents $G^{2}(0)$ for $N_{S} = 0.5$ and $N_{th}$ takes values from $[0, 3.0]$. In this case $G^{2}(0) =1$ at $N_{th} = 0.39$, i.e, this state has zero HBT correlation. The CSTS states for $N_{th} \in [0, 0.39]$, $G^{2}(0) < 1$, i.e., the corresponding states are antibunched.  Graph B indicates a very marginal zero HBT correlated state.   For other fixed values of $N_{S}$, $G^{2}(0)>1$, and the corresponding states are always bunched as shown by graphs C-E.

\section{Conclusion}

The complex tussle and interplay between $N_{S}$ and  $N_{th}$ in the background of a coherent state have compensatory effects which are exhibited by  the photon counting distributions. 
The PCD of a coherent state gets strongly localized by very little addition (of about even $2\%$) of mean squeezed photons; and further, very little addition (of about $2\%)$ of mean thermal photons, remarkably delocalizes the PCD; and, the resulting PCD resembles like that of the initial coherent state. Also, the PCD of a coherent state gets strongly delocalized by very little addition (of about even $2\%$) of mean thermal photons; and further, very little addition (of about $2\%)$ of mean squeezed photons, remarkably relocalizes the PCD; and, the resulting PCD resembles like that of the initial coherent state. This compensatory interplay between the thermal and squeezed photons has interesting effects on the atomic-inversion and entanglement dynamics for both the states SCTS and CSTS. In the case of atomic-inversion for SCTS, a dynamics very similar to the coherent state dynamics is observed for equal contribution of $N_{S}=1$ and $N_{th}=1$ to the coherent state. For $N_{S}=2$ and $N_{th}=2$, the atomic dynamics is  akin to that of a Glauber-Lachs state; i.e., the dominance of thermal photons is evident. The entanglement dynamics $N(t)$ for a SCTS very much resembles that of a Glauber-Lachs state; i.e., the role of $N_{th}$ over $N_{S}$ is again dominant. In the case of CSTS, both $W(t)$ and $N(t)$ show noisy dynamics which can be viewed as the superposition of the dynamics corresponding to a Glauber-Lachs state and a squeezed coherent state. When, $N_{th}$ and $N_{S}$ are increased, both $W(t)$ and $N(t)$ are more under the effects of thermal photons. It is also observed that for both the states SCTS and CSTS, in suitable ranges of $N_{S}$ and $N_{th}$, the states of radiation field show zero HBT correlation and the corresponding states within those ranges are antibunched.

We conclude with a general remark that the tussling interplay between the squeezed photons and the thermal photons has the appearance of an interesting phenomenon in the propagation of an optical soliton in a nonlinear and dispersive medium. The localization and delocalization of the PCD of a coherent state by $N_{S}$ and $N_{th}$ is reminiscent of the interplay between the nonlinearity and dispersion effects in the propagation of optical solitons.

\section*{Acknowledgements}

The authors thank Professors Surendra Singh, S. Sivakumar and  A. B. M. Ahmed for valuable discussions and suggestions.

\bibliographystyle{naturemag}
\bibliography{references.bib}

\begin{thebibliography}{10}
\expandafter\ifx\csname url\endcsname\relax
  \def\url#1{\texttt{#1}}\fi
\expandafter\ifx\csname urlprefix\endcsname\relax\def\urlprefix{URL }\fi
\providecommand{\bibinfo}[2]{#2}
\providecommand{\eprint}[2][]{\url{#2}}

\bibitem{PhysRevA.40.6095}
\bibinfo{author}{Chaturvedi, S.} \& \bibinfo{author}{Srinivasan, V.}
\newblock \emph{\bibinfo{journal}{Phys. Rev. A}} \textbf{\bibinfo{volume}{40}},
  \bibinfo{pages}{6095--6098} (\bibinfo{year}{1989}).
\newblock \urlprefix\url{https://link.aps.org/doi/10.1103/PhysRevA.40.6095}.

\bibitem{PhysRevA.36.1288}
\bibinfo{author}{Janszky, J.} \& \bibinfo{author}{Yushin, Y.}
\newblock \emph{\bibinfo{journal}{Phys. Rev. A}} \textbf{\bibinfo{volume}{36}},
  \bibinfo{pages}{1288--1292} (\bibinfo{year}{1987}).
\newblock \urlprefix\url{https://link.aps.org/doi/10.1103/PhysRevA.36.1288}.

\bibitem{PhysRevA.47.4474}
\bibinfo{author}{Marian, P.} \& \bibinfo{author}{Marian, T.~A.}
\newblock \emph{\bibinfo{journal}{Phys. Rev. A}} \textbf{\bibinfo{volume}{47}},
  \bibinfo{pages}{4474--4486} (\bibinfo{year}{1993}).
\newblock \urlprefix\url{https://link.aps.org/doi/10.1103/PhysRevA.47.4474}.

\bibitem{PhysRevA.47.4487}
\bibinfo{author}{Marian, P.} \& \bibinfo{author}{Marian, T.~A.}
\newblock \emph{\bibinfo{journal}{Phys. Rev. A}} \textbf{\bibinfo{volume}{47}},
  \bibinfo{pages}{4487--4495} (\bibinfo{year}{1993}).
\newblock \urlprefix\url{https://link.aps.org/doi/10.1103/PhysRevA.47.4487}.

\bibitem{PhysRevA.34.3466}
\bibinfo{author}{Vourdas, A.}
\newblock \emph{\bibinfo{journal}{Phys. Rev. A}} \textbf{\bibinfo{volume}{34}},
  \bibinfo{pages}{3466--3469} (\bibinfo{year}{1986}).
\newblock \urlprefix\url{https://link.aps.org/doi/10.1103/PhysRevA.34.3466}.

\bibitem{yi1997squeezed}
\bibinfo{author}{Yi-min, L.}, \bibinfo{author}{Hui-rong, X.},
  \bibinfo{author}{Zu-geng, W.} \& \bibinfo{author}{Zai-xin, X.}
\newblock \emph{\bibinfo{journal}{Acta Physica Sinica (Overseas Edition)}}
  \textbf{\bibinfo{volume}{6}}, \bibinfo{pages}{681} (\bibinfo{year}{1997}).
\newblock \urlprefix\url{https://doi.org/10.1088/1004-423x/6/9/006}.

\bibitem{EZAWA1991216}
\bibinfo{author}{Ezawa, H.}, \bibinfo{author}{Mann, A.},
  \bibinfo{author}{Nakamura, K.} \& \bibinfo{author}{Revzen, M.}
\newblock \emph{\bibinfo{journal}{Annals of Physics}}
  \textbf{\bibinfo{volume}{209}}, \bibinfo{pages}{216--230}
  (\bibinfo{year}{1991}).
\newblock
  \urlprefix\url{https://www.sciencedirect.com/science/article/pii/000349169190360K}.

\bibitem{PhysRevA.40.2494}
\bibinfo{author}{Kim, M.~S.}, \bibinfo{author}{de~Oliveira, F. A.~M.} \&
  \bibinfo{author}{Knight, P.~L.}
\newblock \emph{\bibinfo{journal}{Phys. Rev. A}} \textbf{\bibinfo{volume}{40}},
  \bibinfo{pages}{2494--2503} (\bibinfo{year}{1989}).
\newblock \urlprefix\url{https://link.aps.org/doi/10.1103/PhysRevA.40.2494}.

\bibitem{satyanarayana1992glauber}
\bibinfo{author}{Satyanarayana, M.~V.}, \bibinfo{author}{Vijayakumar, M.} \&
  \bibinfo{author}{Alsing, P.}
\newblock \emph{\bibinfo{journal}{Physical Review A}}
  \textbf{\bibinfo{volume}{45}}, \bibinfo{pages}{5301} (\bibinfo{year}{1992}).
\newblock \urlprefix\url{https://link.aps.org/doi/10.1103/PhysRevA.45.5301}.

\bibitem{subeesh2012effect}
\bibinfo{author}{Subeesh, T.}, \bibinfo{author}{Sudhir, V.},
  \bibinfo{author}{Ahmed, A. B.~M.} \& \bibinfo{author}{Satyanarayana, M.~V.}
\newblock \emph{\bibinfo{journal}{Nonlinear Optics and Quantum Optics}}
  \textbf{\bibinfo{volume}{44}}, \bibinfo{pages}{1--14} (\bibinfo{year}{2012}).
\newblock \urlprefix\url{https://arxiv.org/abs/1203.4792}.

\bibitem{RevModPhys.58.1001}
\bibinfo{author}{Yamamoto, Y.} \& \bibinfo{author}{Haus, H.~A.}
\newblock \bibinfo{title}{Preparation, measurement and information capacity of
  optical quantum states}.
\newblock \emph{\bibinfo{journal}{Rev. Mod. Phys.}}
  \textbf{\bibinfo{volume}{58}}, \bibinfo{pages}{1001--1020}
  (\bibinfo{year}{1986}).
\newblock \urlprefix\url{https://link.aps.org/doi/10.1103/RevModPhys.58.1001}.

\bibitem{PhysRevA.47.5138}
\bibinfo{author}{Kitagawa, M.} \& \bibinfo{author}{Ueda, M.}
\newblock \bibinfo{title}{Squeezed spin states}.
\newblock \emph{\bibinfo{journal}{Phys. Rev. A}} \textbf{\bibinfo{volume}{47}},
  \bibinfo{pages}{5138--5143} (\bibinfo{year}{1993}).
\newblock \urlprefix\url{https://link.aps.org/doi/10.1103/PhysRevA.47.5138}.

\bibitem{PhysRevA.61.010303}
\bibinfo{author}{Ralph, T.~C.}
\newblock \bibinfo{title}{Continuous variable quantum cryptography}.
\newblock \emph{\bibinfo{journal}{Phys. Rev. A}} \textbf{\bibinfo{volume}{61}},
  \bibinfo{pages}{010303} (\bibinfo{year}{1999}).
\newblock \urlprefix\url{https://link.aps.org/doi/10.1103/PhysRevA.61.010303}.

\bibitem{PhysRevA.61.022309}
\bibinfo{author}{Hillery, M.}
\newblock \bibinfo{title}{Quantum cryptography with squeezed states}.
\newblock \emph{\bibinfo{journal}{Phys. Rev. A}} \textbf{\bibinfo{volume}{61}},
  \bibinfo{pages}{022309} (\bibinfo{year}{2000}).
\newblock \urlprefix\url{https://link.aps.org/doi/10.1103/PhysRevA.61.022309}.

\bibitem{PhysRevLett.80.869}
\bibinfo{author}{Braunstein, S.~L.} \& \bibinfo{author}{Kimble, H.~J.}
\newblock \bibinfo{title}{Teleportation of continuous quantum variables}.
\newblock \emph{\bibinfo{journal}{Phys. Rev. Lett.}}
  \textbf{\bibinfo{volume}{80}}, \bibinfo{pages}{869--872}
  (\bibinfo{year}{1998}).
\newblock \urlprefix\url{https://link.aps.org/doi/10.1103/PhysRevLett.80.869}.

\bibitem{PhysRevA.60.937}
\bibinfo{author}{Milburn, G.~J.} \& \bibinfo{author}{Braunstein, S.~L.}
\newblock \bibinfo{title}{Quantum teleportation with squeezed vacuum states}.
\newblock \emph{\bibinfo{journal}{Phys. Rev. A}} \textbf{\bibinfo{volume}{60}},
  \bibinfo{pages}{937--942} (\bibinfo{year}{1999}).
\newblock \urlprefix\url{https://link.aps.org/doi/10.1103/PhysRevA.60.937}.

\bibitem{hu2013statistical}
\bibinfo{author}{Hu, L.-Y.} \& \bibinfo{author}{Zhang, Z.-M.}
\newblock \emph{\bibinfo{journal}{JOSA B}} \textbf{\bibinfo{volume}{30}},
  \bibinfo{pages}{518--529} (\bibinfo{year}{2013}).
\newblock
  \urlprefix\url{https://opg.optica.org/josab/abstract.cfm?URI=josab-30-3-518}.

\bibitem{israel2019entangled}
\bibinfo{author}{Israel, Y.} \emph{et~al.}
\newblock \emph{\bibinfo{journal}{Optica}} \textbf{\bibinfo{volume}{6}},
  \bibinfo{pages}{753--757} (\bibinfo{year}{2019}).
\newblock
  \urlprefix\url{https://opg.optica.org/optica/abstract.cfm?URI=optica-6-6-753}.

\bibitem{photonics8030072}
\bibinfo{author}{Mouloudakis, G.} \& \bibinfo{author}{Lambropoulos, P.}
\newblock \emph{\bibinfo{journal}{Photonics}} \textbf{\bibinfo{volume}{8}}
  (\bibinfo{year}{2021}).
\newblock \urlprefix\url{https://www.mdpi.com/2304-6732/8/3/72}.

\bibitem{simidzija2018harvesting}
\bibinfo{author}{Simidzija, P.} \& \bibinfo{author}{Martin-Martinez, E.}
\newblock \emph{\bibinfo{journal}{Physical Review D}}
  \textbf{\bibinfo{volume}{98}}, \bibinfo{pages}{085007}
  (\bibinfo{year}{2018}).
\newblock \urlprefix\url{https://link.aps.org/doi/10.1103/PhysRevD.98.085007}.

\bibitem{wang2017statistical}
\bibinfo{author}{Wang, Z.}, \bibinfo{author}{Li, H.-M.}, \bibinfo{author}{Yuan,
  H.-C.}, \bibinfo{author}{Wan, Z.-L.} \& \bibinfo{author}{Meng, X.-G.}
\newblock \emph{\bibinfo{journal}{International Journal of Theoretical
  Physics}} \textbf{\bibinfo{volume}{56}}, \bibinfo{pages}{729--740}
  (\bibinfo{year}{2017}).
\newblock \urlprefix\url{https://doi.org/10.1007/s10773-016-3214-5}.

\bibitem{Dupays2021shortcutstosqueezed}
\bibinfo{author}{Dupays, L.} \& \bibinfo{author}{Chenu, A.}
\newblock \emph{\bibinfo{journal}{{Quantum}}} \textbf{\bibinfo{volume}{5}},
  \bibinfo{pages}{449} (\bibinfo{year}{2021}).
\newblock \urlprefix\url{https://doi.org/10.22331/q-2021-05-01-449}.

\bibitem{PhysRevLett.122.040602}
\bibinfo{author}{Klaers, J.}
\newblock \emph{\bibinfo{journal}{Phys. Rev. Lett.}}
  \textbf{\bibinfo{volume}{122}}, \bibinfo{pages}{040602}
  (\bibinfo{year}{2019}).
\newblock
  \urlprefix\url{https://link.aps.org/doi/10.1103/PhysRevLett.122.040602}.

\bibitem{PhysRev.131.2766}
\bibinfo{author}{Glauber, R.~J.}
\newblock \emph{\bibinfo{journal}{Phys. Rev.}} \textbf{\bibinfo{volume}{131}},
  \bibinfo{pages}{2766--2788} (\bibinfo{year}{1963}).
\newblock \urlprefix\url{https://link.aps.org/doi/10.1103/PhysRev.131.2766}.

\bibitem{satyanarayana1989ringing}
\bibinfo{author}{Satyanarayana, M.~V.}, \bibinfo{author}{Rice, P.},
  \bibinfo{author}{Vyas, R.} \& \bibinfo{author}{Carmichael, H.}
\newblock \emph{\bibinfo{journal}{JOSA B}} \textbf{\bibinfo{volume}{6}},
  \bibinfo{pages}{228--237} (\bibinfo{year}{1989}).
\newblock \urlprefix\url{https://doi.org/10.1364/JOSAB.6.000228}.

\bibitem{jaynes1963comparison}
\bibinfo{author}{Jaynes, E.~T.} \& \bibinfo{author}{Cummings, F.~W.}
\newblock \emph{\bibinfo{journal}{Proceedings of the IEEE}}
  \textbf{\bibinfo{volume}{51}}, \bibinfo{pages}{89--109}
  (\bibinfo{year}{1963}).
\newblock \urlprefix\url{https://doi.org/10.1109/PROC.1963.1664}.

\bibitem{gerry2005introductory}
\bibinfo{author}{Gerry, C.}, \bibinfo{author}{Knight, P.} \&
  \bibinfo{author}{Knight, P.~L.}
\newblock \emph{\bibinfo{title}{Introductory Quantum Optics}}
  (\bibinfo{publisher}{Cambridge University Press}, \bibinfo{year}{2005}).

\bibitem{wei2003maximal}
\bibinfo{author}{Wei, T.-C.} \emph{et~al.}
\newblock \emph{\bibinfo{journal}{Physical Review A}}
  \textbf{\bibinfo{volume}{67}}, \bibinfo{pages}{022110}
  (\bibinfo{year}{2003}).
\newblock \urlprefix\url{https://link.aps.org/doi/10.1103/PhysRevA.67.022110}.

\end{thebibliography}

\end{document}